\documentclass[twocolumn]{aastex631}
\usepackage[utf8]{inputenc}
\usepackage{amsmath,amsxtra,amssymb,latexsym,amscd,amsthm}
\usepackage{mathrsfs}
\usepackage[mathscr]{eucal}
\usepackage{mathrsfs}
\usepackage{makeidx}
\usepackage{soul}
\usepackage{graphicx}
\usepackage{subfigure}
\usepackage{booktabs}
\usepackage{multirow}
\usepackage{threeparttable}
\usepackage{hyperref}

\shorttitle{Dust Dynamics in VSI Turbulence}
\shortauthors{Huang \& Bai}
\graphicspath{{./}{figures/}}

\begin{document}

\title{The Interplay between Dust Dynamics and Turbulence Induced by the Vertical Shear Instability}

\author[0000-0002-7575-3176]{Pinghui Huang \dag}
\affiliation{CAS Key Laboratory of Planetary Sciences, Purple Mountain Observatory, Chinese Academy of Sciences, Nanjing 210008, People’s Republic of China}
\affiliation{Department of Physics \& Astronomy, University of Victoria, Victoria, British Columbia, V8P 5C2, Canada}
\affiliation{Institute for Advanced Study, Tsinghua University, Beijing 100084, People’s Republic of China}

\author[0000-0001-6906-9549]{Xue-Ning Bai \ddag}
\affiliation{Institute for Advanced Study, Tsinghua University, Beijing 100084, People’s Republic of China}
\affiliation{Department of Astronomy, Tsinghua University, Beijing 100084, People’s Republic of China}

\correspondingauthor{Pinghui Huang; Xue-Ning Bai}
\email{phhuang@pmo.ac.cn; xbai@tsinghua.edu.cn}

\begin{abstract}
The interaction between gas and dust in protoplanetary disks (PPDs) plays a crucial role in setting the stage of planet formation. In particular, the streaming instability (SI) is well recognized as the mechanism for planetesimal formation out of this interaction. The outer region of PPDs is likely subject to the vertical shear instability (VSI), representing a major source of disk turbulence characterized by vertical corrugation that leads to strong dust stirring. In the meantime, the VSI turbulence in 3D generates vortices through the Rossby wave instability (RWI), which can trap dust and thereby promote dust concentration. In this study, we use the multifluid dust module in Athena++ to conduct 2D axisymmetric global simulations of PPDs with mesh refinement and 3D global simulations with modest resolution. In 2D, the VSI corrugation mode is weakened by dust back-reaction, while the SI can still survive regardless of initial conditions. Dust clumping occurs and is seeded by VSI-induced zonal flows. In 3D, dust can settle even more with increased dusty buoyancy, suppressing the VSI corrugation mode. Meanwhile, dust back-reaction enhances dust concentration in RWI vortices, though higher resolution is needed to assess dust clumping.
\end{abstract}

\section{Introduction}~\label{sec:intro}

Protoplanetary disks (PPDs) are gaseous disks surrounding newly-formed stars. With typical lifetime of a few million years (Myrs)~\citep{WilliamsCieza2011,Andrews2020}, PPDs set the stage for planet formation, where dust grains in PPDs grow from sub-micron size all the way by more than 40 orders of magnitude in mass through a series of physical processes~\citep{Armitage2020}. Crucial to many aspects of planet formation are the gas dynamics in PPDs. These include the mechanisms that governs disk angular momentum transport, which determines the disk structure and drives disk evolution, and the level of turbulence, which strongly influences most processes of dust growth towards planets. In this work, we aim at studying one of the most uncertain aspect of planet formation, dust clumping towards planetesimal formation, under the influence turbulence generated by the vertical shear instability (VSI,~\citealp{Nelson2013}), applicable to a wide range of radii in PPDs.

The physics governing the gas dynamics in PPDs are characterized by two facts. First, the disk is poorly ionized so that the coupling with magnetic field is weak. This leads to three non-ideal magnetohydrodynamic (MHD) effects that largely suppress the turbulence generated by the magnetorotational instability (MRI,~\citealp{BalbusHawley1998}) in the inner $\sim10$AU region~\citep{Gammie1996,BaiStone2013Wind,Gressel2015}. The MRI turbulence in the outer PPDs is likely damped~\citep{BaiStone2011,Simon2013,CuiBai2021}, maintaining at relatively low level. On the other hand, angular momentum transport is likely driven primarily by a magnetized disk wind~\citep{BlandfordPayne1982,BaiStone2017}. Second, under the condition that the MRI turbulence is highly suppressed or damped, several hydrodynamic instabilities have recently been identified that can operate depending on the disk thermodynamic conditions~\citep{FromangLesur2019,LyraUmurhan2019,Lesur2023}. These include the VSI~\citep{UrpinBrandenburg1998,Urpin2003,ArltUrpin2004,Nelson2013,BarkerLatter2015}, the Convective Overstability~\cite[COS,][]{KlahrBodenheimer2003,Lyra2014,Latter2016}, and the Zombie Vortex Instability~\cite[ZVI,][]{BarrancoMarcus2005,Marcus2015,Marcus2016}. In particular, the VSI is an axisymmetric instability that originates from the baroclinic term in protoplanetary disks~\citep{GoldreichSchubert1967,Fricke1968}. It operates under rapid cooling conditions that are expected to be generically satisfied in the outer regions of PPDs~\cite[beyond $\sim10$ AU, ][]{LinYoudin2015,PfeilKlahr2019}. Its nonlinear development generates moderate, anisotropic turbulence, influencing disk evolution~\citep{Nelson2013,BarkerLatter2015}, and can efficiently stir up dust in the disk~\citep{StollKley2014,StollKley2016,Dullemond2022}. Over the past 10 years, substantial amount of works have been devoted to understanding the nature and properties of the VSI, including its linear onset, physical origin, turbulence properties, and three-dimensional effects.  Encouragingly, the VSI turbulence remains robust under magnetized disk wind~\citep{CuiBai2020} and dampened MRI turbulence~\citep{CuiBai2022}.

Dust growth, transport and planetesimal formation represent the initial stage of planet formation~\citep{BirnstielDullemond2010,Birnstiel2023}. The dust particles are coupled with the gas via aerodynamic drag. Being pressureless, they naturally drift toward gas pressure maxima~\citep{Weidenschilling1977Aerodynamics,Weidenschilling1980Planetesimals,BargeSommeria1995}, and diffuse in response to disk turbulence~\citep{YoudinLithwick2007}. It has been widely recognized that planetesimal formation proceeds through certain mechanisms that first concentrate dust into dense clumps, followed by gravitational collapse over their own self-gravity~\citep{ChiangYoudin2010}. The streaming instability~\cite[SI,][]{YoudinGoodman2005} has been considered to be the most promising mechanism for planetesimal formation. It arises from the radial drift of dust against a radial pressure gradient in gaseous disks, and is two-fluid in nature requiring dust back-reaction to operate, which can also be considered as a flavor of the resonant drag instability~\citep{HopkinsSquire18}. Simulations of the SI, mostly under the local shearing box framework, indicate that the SI can lead to strong dust clumping to sufficiently high densities and trigger gravitational collapse to directly form planetesimals~\citep{JohansenOishi2007,SimonArmitage2017}. The condition for sufficient dust clumping generally requires reaching a threshold dust abundance~\citep{Johansen2009,LiYoudin2021}, whose value depend on dust size distribution~\citep{Bai2010dynamics,CarreraJohansen2015,YangJohansen2017}, radial pressure gradient~\citep{Bai2010pressure,Abod2019}.

Most studies of the SI have assumed that the disk background is laminar without external turbulence. However, it has been found that external disk turbulence, when treated as viscosity, can suppress the SI~\citep{Umurhan2020,ChenLin2020}. The same conclusion holds when the system simulated with artificially-driven hydrodynamic turbulence~\citep{GoleSimno2020,LimSimon2024}. Therefore, it is natural to ask about the fate of the SI under {\it realistic} turbulence in PPDs. Recent studies have suggested that in the case of the MRI turbulence, while dust clumping can be observed, it occurs only at radial pressure maxima, a condition that is inconsistent with that for the SI to operate~\citep{XuBai2022}. On the other hand,~\citet{SchaferJohansen2020,SchaferJohansen2022,Schafer2025} recently found that the SI is able to coexist with VSI turbulence in 2D simulation, but the condition for dust clumping is substantially modified. They also reported that the interplay between the SI and VSI appears to depend on initial condition regarding which instability was first activated, which we will revisit in this work.

The behavior of the VSI turbulence differ substantially in 3D: it can generate zonal flows that lead to the formation of Rossby-wave instability (RWI) vortices~\citep{Richard2016,MangerKlahr2018}. As vortices themselves are naturally dust traps with pressure maxima in the center, they very likely lead to an alternative path to planetesimal formation, as has been studied in other contexts~\citep{Raettig2015,Raettig2021}. Even without sufficient resolution to resolve the SI, the mass loading by the dust is already sufficient to alter the properties of the VSI turbulence as found in 2D studies~\citep{Lin2019}, and we expect a similar influence should carry over to 3D.

In this paper, we utilize the newly developed multifluid dust module~\citep{HuangBai2022} in the MHD code Athena++~\citep{Stone2020} to investigate the interplay between dust dynamics and the VSI turbulence in both 2D and 3D. Our 2D simulations revisit the recent study of~\citet{SchaferJohansen2020,SchaferJohansen2022,Schafer2025} who used particle treatment of dust. Our work validates the fluid treatment of dust, and features the use of mesh refinement to achieve higher resolution and larger computational domain, demonstrating that the outcome is independent of the initial condition. We then extend the simulations to 3D for the first time,
focusing on how dust back-reaction influences general gas dynamics without including mesh refinement. A follow-up paper will enable mesh refinement, dedicated to understanding dust clumping and planetesimal formation from the 3D simulations. The outline of this paper is as follows. Section~\ref{sec:method} describes the governing equations, numerical methods, and diagnostics employed. We present in Section~\ref{sec:2DRunsVSIandSI} and Section~\ref{subsec:3DRunsVSIandRWI} the 2D and 3D numerical simulation results, progressing with increasing complexity. Finally in Section~\ref{sec:summary}, we discuss the implications of our findings and conclude.

\section{Simulation Setup and Diagnostics}~\label{sec:method}

\subsection{Formulation and Problem Setup}~\label{subsec:equationsAndsetup}

We conduct global simulations of gas and dust using the multifluid dust module in the Athena++ MHD code~\citep{Stone2020,HuangBai2022}. The dust is treated as a single pressureless fluid, characterized by the stopping time $T_s$, non-dimensionized by the Stokes number $St\equiv\Omega_K T_s$, where $\Omega_K$ is the Keplerian angular velocity. This treatment is generally valid as long as $St\ll1$, and can well capture the non-linear evolution of the SI with dust clumping~\citep{Benitez2019FARGO3D,HuangBai2022}. As we explicitly capture the relevant instabilities and turbulence in disks, we do not incorporate any explicit dissipation such as viscosity and dust diffusion. The general equations for the gas and the dust fluid read
\begin{equation}
\frac{\partial \rho_{\text{g}}}{\partial t} + \nabla \cdot \left(\rho_{\text{g}} \boldsymbol{v}_{\text{g}}\right) =0\ ,
\label{eq:gas_con}
\end{equation}
\begin{equation}
\begin{aligned}
  \frac{\partial \rho_{\text{g}}\boldsymbol{v}_{\text{g}}}{\partial t} +  \nabla\cdot\left(\rho_{\text{g}}\boldsymbol{v}_{\text{g}} \boldsymbol{v}_{\text{g}} + P\mathsf{I}\right)= \rho_{\text{g}} \nabla \Phi +\rho_{\text{d}} \frac{\boldsymbol{v}_{\text{d}} - \boldsymbol{v}_{\text{g}}}{T_{\text{s}}}\ ,
\end{aligned}
\label{eq:gas_mom}
\end{equation}
\begin{equation}
  \frac{\partial \rho_{\text{d}}}{\partial t} +\nabla \cdot \left(\rho_{\text{d}} \boldsymbol{v}_{\text{d}}\right) = 0\ ,
\label{eq:dust_con}
\end{equation}
\begin{equation}
\begin{aligned}
  \frac{\partial \rho_{\text{d}} \boldsymbol{v}_{\text{d}}}{\partial t} + \nabla\cdot\left(\rho_{\text{d}}\boldsymbol{v}_{\text{d}} \boldsymbol{v}_{\text{d}} \right) = \rho_{\text{d}} \nabla \Phi  + \rho_{\text{d}}\frac{\boldsymbol{v}_{\text{g}} - \boldsymbol{v}_{\text{d}}}{T_{\text{s}}}\ ,
\end{aligned}
\label{eq:dust_mom}
\end{equation}
where the subscripts ``g'' and ``d'' denote gas and dust, respectively, $\rho$, $P$, $\mathbf{v}$ denote density, pressure and velocity, $\mathsf{I}$ represents the identity matrix, and $\nabla \Phi$ corresponds to central stellar gravity, i.e., $\Phi = GM/r$, where $GM = 1$ represents the gravitational constant multiplied by the stellar mass. Unless specified otherwise, we apply a locally isothermal equation of state:
\begin{equation}
P = c_\text{s}^2 (R) \rho_\text{g}.
\label{eq:gas_pre}
\end{equation}
where $c_\text{s}$ is the isothermal sound speed. This means that the effective cooling time is zero, which can efficiently trigger the VSI. We ignore the effects from the self-gravity and the magnetic field in this study, leave them in the future works.

The simulations are conducted in spherical-polar coordinates $(r, \theta, \phi)$, while results are typically analyzed in cylindrical coordinates $(R, \phi, z)$.  In all simulations, the radial range is set to $r \in [1, 3]$ with logarithmic spacing, $\theta$ spans from $\pi/2 - 0.4$ to $\pi/2+0.4$ with uniform spacing. In the 3D simulations, the azimuthal domain, $\phi$ covers the range from 0 to $\pi$ with uniform spacing, which is found to be sufficient to properly capture the saturation of the VSI turbulence~\citep{MangerKlahr2018}. At root level, the grid size in $r-\theta$ is set to $512\times 256$ in 2D, 768 in $\phi$ in 3D simulations, achieving an effective resolution about $40 \times 28 \times 22 $ cells per $H_\text{g}$ along $r$, $\theta$ and $\phi$ directions ($\Delta r$:$r \Delta \theta$:$r \Delta \phi$ = $1:1.4:1.8$) at $R = 1.5$.

\subsection{Initial and Boundary Conditions}~\label{subsec:ICandBC}

In this work, we consider the gas density at the midplane and the isothermal sound speed following a power-law profile with respect to the cylindrical radius $R$: $\rho_\text{g,mid}(R) = \rho_\text{g,0} \left(R/R_0 \right)^p$ and $c_\text{s}^2(R) = c_\text{s,0}^2 \left(R/R_0 \right)^q$, where $\rho_\text{g,0} = 1$ and $c_\text{s,0} = 0.08$ at $R_0 \equiv 1$ throughout this paper. The power law indices are taken to be $p = -3/2$ and $q = -1/2$. The initial gas density $\rho_\text{g}$ and the gas orbital frequency $\Omega_\text{g}$ are given by~\citep{Nelson2013}
\begin{equation}
  \rho_\text{g,init}(R,z) = \rho_\text{g,mid} \exp \left[\frac{GM}{c_\text{s}^2}\left(\frac{1}{r}-\frac{1}{R}\right) \right],
\label{eq:init_density}
\end{equation}
\begin{equation}
  \Omega_\text{g,init}(R,z) = \Omega_\text{K} \left[\left(p+q\right)h^2 + \left(1+q\right) - q\frac{R}{r}\right]^{\frac{1}{2}}.
\label{eq:init_omega}
\end{equation}
where the Keplerian frequency is $\Omega_\text{K}(R) \equiv \sqrt{GM/R^3}$, disk aspect ratio is $h(R) \equiv  H_\text{g}/R$ and gas scale height is $H_\text{g}(R) \equiv c_\text{s}/\Omega_\text{K}$. With our choice of $p$ and $q$, the disk is flared with $h \propto R^{\frac{1}{4}}$, and the disk surface density $\Sigma\propto R^{-1/4}$. Based on~\citet{Umurhan2016}, the VSI body modes with the fastest growth rates have radial wavelengths of approximately $\lambda_\text{fast} \approx \pi h H_\text{g} = 0.0064 \pi \left( \frac{R}{R_0} \right)^{3/2} R_0$, which are resolved by about 10 to 20 grids at the root grid level in our simulations. We note that our $z-$domain covers $\pm 5H_\text{g}$ at $R = 1$ and $\pm 3.5 H_\text{g}$ at $R = 2.7$, which is much larger than that of~\cite{SchaferJohansen2020} who took $\lesssim \pm 2H_\text{g}$. The initial gas velocities are set as: $v_{\text{g},r,\text{init}} = v_{\text{g},R,\text{init}} = 0$, $v_{\text{g},\theta,\text{init}} = v_{\text{g},z,\text{init}} = 0$ and $v_{\text{g},\phi,\text{init}} = R \Omega_{\text{g,init}} $, and we further add $\sim 0.01 \;c_\text{s}$ white noise perturbations on the gas velocities to seed the instabilities.

The initial dust density is set as $\rho_\text{d,init} = Z_\text{d} \rho_\text{g,init}$, with $Z_\text{d} = 0.01$. The initial cylindrical radial, azimuthal and vertical dust velocities are determined using the Nakagawa-Sekiya-Hayashi (NSH) solution~\citep{NakagawaSekiyaHayashi1986}:
\begin{equation}
  v_{\text{d},R,\text{init}} = \frac{2}{St + \left(1+Z_\text{d}\right)^2 St^{-1}}\eta v_\text{K},
\end{equation}
\begin{equation}
  v_{\text{d},\phi,\text{init}} = \left[ 1 + \frac{\left(1+Z_\text{d}\right)}{\left(1+Z_\text{d}\right)^2+ St^{2}} \right] \eta v_\text{K},
\end{equation}
\begin{equation}
  v_{\text{d},z,\text{init}} = 0.
\end{equation}
where
$\eta \equiv (1/2)(c_\text{s}/v_\text{K})^2 (d \log{P}/d \log{R})$
represents the strength of radial pressure gradient. For simplicity, we maintain a fixed Stokes number of dust throughout this study, i.e., $St = 0.1$, which typically correspond to (sub-)millimeter-sized dust located within tens to hundreds of AU in protoplanetary disks.

We set the $r$ and $\theta$ boundary conditions by fixing the values at the ghost cells to initial state, while the $\phi-$boundary condition is simply periodic. We use the Harten-Lax-Van Leer (HLLC) Riemann solver for gas and the default Riemann solver described by~\cite{HuangBai2022} for dust, the Piecewise Linear Method (PLM) spatial reconstruction and the van-Leer (VL2) time integrator with Courant-Friedrichs-Lewy (CFL) number set to 0.3 in all 2D and 3D simulations. We also use orbital advection~\citep{Masset2000,Masset2002,Stone2020} on the azimuthal velocities for both gas and dust in all simulations, so as to speed up the calculations.

We set the density floors for gas and dust to stabilize the simulations: $\rho_\text{g,floor} = \max(10^{-6}, 10^{-2}\rho_\text{g,init})$ and $\rho_\text{d,floor} = \max(10^{-8}, 2\times 10^{-3}\rho_\text{d,init})$, where $\rho_\text{g,init}$ and $\rho_\text{d,init}$ are the initial densities for gas and dust at corresponding locations.

\begin{table*}
\begin{normalsize}
\caption{Problem Setup for 2D and 3D Runs}
\begin{tabular}{ccccc}
\toprule
\hline
Model Name            & Mesh Refinement  & Dust Release\tnote{$^1$}   & Equation of State\tnote{$^2$}  & Dust Feedback\tnote{$^3$} \\
\hline
2D-Only SI           & At time 0  & At time 0    & Adiabatic          & Yes           \\
\hline
2D-Only VSI           & At time 0  & At time 0    & Locally Isothermal          & No          \\
\hline
2D-SIAfterVSI       & After 300 Orbits & After 300 Orbits & Locally Isothermal & Yes           \\
\hline
2D-SIWhileVSI       & At time 0  & At time 0    & Locally Isothermal & Yes           \\
\hline
3D-NoFB (3D-VSI\&RWI) & No               & At time 0    & Locally Isothermal & No            \\
\hline
3D-FB                 & No               & At time 0    & Locally Isothermal & Yes           \\
\bottomrule
\end{tabular}
\label{tab:Models}
\end{normalsize}
\end{table*}

\subsection{Numerical Diagnostics}~\label{subsec:diagnostics}

To investigate the properties of gas turbulence and the interaction between dust and gas, here we list several numerical diagnostics focusing on the VSI turbulence, SI-induced dust clumping, and the RWI vortices.

Upon the development of the VSI turbulence, we quantify the strength of the turbulence by the radial, azimuthal and vertical Mach numbers (of the gas), defined as:
\begin{equation}
  Ma_R \equiv v_{\text{g},R}/c_\text{s},\quad
  Ma_\phi \equiv \delta v_{\text{g},\phi}/c_\text{s},\quad
  Ma_z \equiv v_{\text{g},z}/c_\text{s}\ .
\label{eq:Ma}
\end{equation}
where $\delta v_{\text{g},\phi}$ is the azimuthal velocity difference compared to the initial azimuthal gas velocity, $\delta v_{\text{g},\phi} \equiv v_{\text{g},\phi} - v_{\text{g},\phi,\text{init}}$. A more quantitative diagnostics of turbulence is the kinetic energy power spectrum. We can calculate the total kinetic energy variation as:
\begin{equation}
\begin{aligned}
  & \delta E_\text{kin}  \equiv \frac{1}{2}  \rho_\text{g} \left[v_{\text{g},R}^2 + \delta v_{\text{g},\phi}^2+ v_{\text{g},z}^2 \right] + \\  & \frac{1}{2} \rho_\text{d} \left[\left(v_{\text{d},R} - v_{\text{d},R,\text{init}}\right)^2 + \left(v_{\text{d},\phi} - v _{\text{d},\phi,\text{init}}\right)^2 + v_{\text{d},z}^2\right].
\end{aligned}
\label{eq:kinerg}
\end{equation}
We can then carry out fast Fourier transform to calculate the power spectrum of $\delta E_\text{kin}$ about the azimuthal wavenumber $k_\phi$ at the midplane ($\theta = \pi/2, z = 0$):
\begin{equation}
\begin{aligned}
\mathcal{F} (k_\phi) = \int^{2\pi}_0 \delta E_\text{kin,midplane}\;e^{-2 \pi i k_\phi R\phi}\; R d\phi.
\end{aligned}
\label{eq:fft_kinerg}
\end{equation}
Besides, the effective turbulent viscosity associated with radial transport of angular momentum $\alpha$ is defined as:
\begin{equation}
  \alpha_{R\phi} \equiv \langle |\frac{\rho_\text{g} v_{\text{g},R}\delta v_{\text{g},\phi}}{\rho_\text{g,init} c_\text{s}^2}|\rangle_\phi.
\label{eq:alpha_Rphi}
\end{equation}
where $\langle \;\rangle_\phi$ means the azimuthal averaging.

In 3D, the VSI turbulence generates radial density/pressure bumps that are subject to the Rossby wave instability~\citep[RWI,][]{LovelaceLi1999,LiFinn2000,Ono2016}. More quantitatively, one can define a function named ``Entropy Modified Inversed Vortensity'',
\begin{equation}
  \mathscr{L} \equiv  \frac{\Sigma_\text{g} s^{2/\gamma}}{2 \left(\vec{\omega} \cdot \hat{z}\right)}\ ,
\label{eq:Lfunc}
\end{equation}
where $s$ is the specific gas entropy. The RWI occurs where $\mathscr{L}$ reaches local maximum. In the locally isothermal case, the function $\mathscr{L}$ can be further reduced into a simpler form at the midplane case~\citep{Lin2012a}:
\begin{equation}
  \mathscr{L}_\text{iso,midplane} \equiv \frac{\langle \Sigma_\text{g}\rangle_\phi \langle c_\text{s,iso}^2\rangle_\phi }{\langle \omega_{z,\text{midplane}}\rangle_\phi}\ .
\label{eq:Lfunc_iso}
\end{equation}
where $\vec{\omega}$ is the gas vorticity.

The onset of the RWI can yield numerous vortices. The properties of the vortices can be characterized by vertical vorticity compared to the background shear
\begin{equation}
  \omega_z \equiv \left[\nabla \times \left(\mathbf{v_\text{g}} - \mathbf{v_\text{g,init}}\right) \right]_z\ .
\label{eq:vorticity}
\end{equation}
Another important diagnostics for vortices is their aspect ratio:
\begin{equation}
\chi \equiv r_\text{c} \Delta \phi / \Delta r\ .
\label{eq:aspect}
\end{equation}
where $r_\text{c}$ is the radial location of the vortex center, $\Delta \phi$  $\Delta r$ are its azimuthal and radial width.

The effective entropy $S_\text{eff}$ and squared vertical dusty buoyancy frequency $N_\text{z,eff}^2$ modified by dust back-reaction are defined by~\cite{Lin2019}:
\begin{equation}
  S_\text{eff} \equiv \ln{\frac{P^{1/\gamma}}{\rho_\text{eff}}},
\label{eq:entropy}
\end{equation}
\begin{equation}
  N_\text{z,eff}^2 \equiv -\frac{1}{\rho_\text{eff}}\frac{\partial P}{\partial z} \frac{\partial S_\text{eff}}{\partial z}.
\label{eq:buoyancy}
\end{equation}
where $\gamma = 1$ for the locally isothermal disks. In the above, the effective density of the dust-gas mixture can be simply taken as $\rho_\text{eff} \equiv  \rho_\text{d} + \rho_\text{g}$~\citep{Lin2019}, which largely identical to the more rigorous approach in~\citet{Bai2010dynamics} for $St\lesssim0.1$.

In the saturated state, dust settling balances turbulent diffusion and the dust layer sustains a relatively constant thickness. We can calculate the dust scale height as:
\begin{equation}
  H_\text{d} \equiv \sqrt{\frac{\int^{z_\text{max}}_{z_\text{min}} \langle \rho_\text{d} z^2\rangle_\phi dz}{\int^{z_\text{max}}_{z_\text{min}} \langle \rho_\text{d}\rangle_\phi dz}},
\label{eq:dust_scaleheight}
\end{equation}
Over the range of interest, we can define the radially-averaged dust-gas scale height ratio as
\begin{equation}
	\langle \frac{H_\text{d}}{H_\text{g}}\rangle_R \equiv \frac{\int^{R_\text{max}}_{R_\text{min}} \frac{H_\text{d}}{H_\text{g}} dR}{\int^{R_\text{max}}_{R_\text{min}} dR}.
\label{eq:averaging_Hd}
\end{equation}
By default, we calculate the radially-averaged value between $R_\text{min} = 1.2$ and $R_\text{max} = 2.4$.

Our simulations do not include self-gravity and hence do not directly follow the gravitational collapse of dust clumps following dust clumping. On the other hand, we can estimate the Roche density $\rho_\text{R}$~\cite[e.g.][]{Bai2010dynamics,LiYoudin2021}:
\begin{equation}
  \rho_\text{R} \equiv  \frac{9}{4}\sqrt{2 \pi} Q \rho_\text{g},
\label{eq:Roche}
\end{equation}
where $Q\equiv c_\text{s} \Omega_\text{K}/(\pi G \Sigma_\text{g})$, is the Toomre $Q$ parameter~\citep{Toomre1964}. Using the typical value of $Q = 32$ at $R_0 = 1$ in the outer part of Class II disks, the Roche density is $\rho_\text{R} = 180 \left(\frac{R}{R_0}\right)^{-3/2}$ in this study.
The dust clump becomes self-gravitating, leading to planetesimal formation when $\rho_d>\rho_R$. For practical purpose, we adopt a critical dust-gas density ratio taken at $R=1.5$:
\begin{equation}
\epsilon_\text{crit} \equiv \left(\rho_\text{d}/\rho_\text{g}\right)_\text{crit} = 98.
\label{eq:crit_ratio}
\end{equation}
as the criteria of planetesimal formation.

In Appendix~\ref{app:symbols}, we list all symbols used in this study.

\subsection[]{List of simulations}

We conduct four 2D axisymmetric runs and two 3D runs, and their basic setups are listed in Table~\ref{tab:Models}. More specifically, for 2D simulations, we apply three levels of mesh refinement in the radial domain $r \in [1.2, 2.4]$ and the meridional domain $\theta \in [\pi/2 - 0.056, \pi/2 + 0.056]$ (about $318 \times 227$ cells per $H_\text{g}^2$ along the $r-\theta$ directions at $R = 1.5$ and $z = 0$). The simulations are labeled as ``2D-Only SI'', ``2D-Only VSI'', ``2D-SIAfterVSI'' and ``2D-SIWhileVSI''.

In the ``2D-Only SI'' case, we suppress the VSI by employing the adiabatic equation of state. This is not expected to affect the SI, as its growth rates are comparable in adiabatic and isothermal regimes~\citep[see Figure 8 in ][]{LehmannLin2023}. In doing so, an additional gas energy equation is included ~\cite[see Equation 3 in ][]{HuangBai2022}:
\begin{equation}
\begin{aligned}
\frac{\partial E_{\text{g}}}{\partial t} +& \nabla \cdot\left[\left(E_{\text{g}}+P\right) \boldsymbol{v}_{\text{g}}\right]= \rho_{\text{g}} \nabla\Phi \cdot \boldsymbol{v}_{\text{g}} \\
&+ \rho_{\text{d}} \frac{\boldsymbol{v}_{\text{d}} - \boldsymbol{v}_{\text{g}}}{T_{\text{s}}}\cdot \boldsymbol{v}_{\text{g}} + \rho_{\text{d}} \frac{\left(\boldsymbol{v}_{\text{d}} - \boldsymbol{v}_{\text{g}}\right)^2}{T_{\text{s}}}
\end{aligned}
\label{eq:gas_erg}
\end{equation}
Note that SI turbulence and friction with dust lead to additional heating. However, given the relatively low turbulence level, the resulting increase in gas temperature is largely negligible, with a relative difference of less than 5\% over the duration of our simulations. This run serves as a baseline for assessing the role of VSI on SI. Additionally, we conduct the ``2D-Only VSI'' model using a locally isothermal equation of state (Equation~\ref{eq:gas_pre}), where dust feedback (the second term on the right-hand side of Equation~\ref{eq:gas_mom} and~\ref{eq:dust_mom}) is turned off to suppress the occurrence of SI, providing another baseline test.

In subsequent 2D VSI + SI simulations, following~\cite{SchaferJohansen2020},~\cite{SchaferJohansen2022}, we consider two scenarios: whether SI is triggered simultaneously with or after the onset of VSI turbulence. In the ``2D-SIWhileVSI'' simulations, dust is released initially, whereas in the ``2D-SIAfterVSI'' simulations, dust is released after 300 innermost orbits (at $R = R_0$). By this time, we observe that the VSI is fully developed across the domain, as indicated by the complete mixing of gas, tracked using passive scalar tracers.

Mesh refinement is activated initially in the ``2D-Only SI'', ``2D-Only VSI'' and ``2D-SIWhileVSI'' models, whereas it is activated after 300 orbits in the ``2D-SIAfterVSI'' models. To ensure robustness, we conducted additional tests to assess the timing of mesh refinement (at time 0 or after 300 orbits) across these four models. The results indicate that the timing of mesh refinement has no significant effect on the final turbulent environment in any of the 2D models.

For 3D VSI simulations, we perform two runs: one excluding dust feedback (``3D-NoFB'') and another incorporating dust feedback (``3D-FB''). Since 3D gaseous VSI turbulence is saturated by RWI vortices, the ``3D-NoFB'' model can also be referred to as the ``3D-VSI\&RWI'' model. Our 3D simulations do not include mesh refinement, resulting in an effective resolution of approximately $33^2$ cells per $H_\text{g}^2$ at $R = 1.5$. This resolution is insufficient to resolve SI in the ``3D-FB'' model, as our primary objective is to investigate the impact of dust feedback on 3D VSI (RWI) turbulence in the ``3D-FB'' model. Simulations incorporating mesh refinement to resolve the SI will be presented in a follow-up publication.

\section{2D Axisymmetric Simulations Results}~\label{sec:2DRunsVSIandSI}
In this section, we present the results from four 2D axisymmetric runs, and the findings are to be compared with standard SI simulations in the literature, and those of~\cite{SchaferJohansen2020}, ~\cite{SchaferJohansen2022} and~\cite{Schafer2025} on the interplay between the SI and VSI.

\begin{figure*}[htp]
\centering
\includegraphics[scale=0.45]{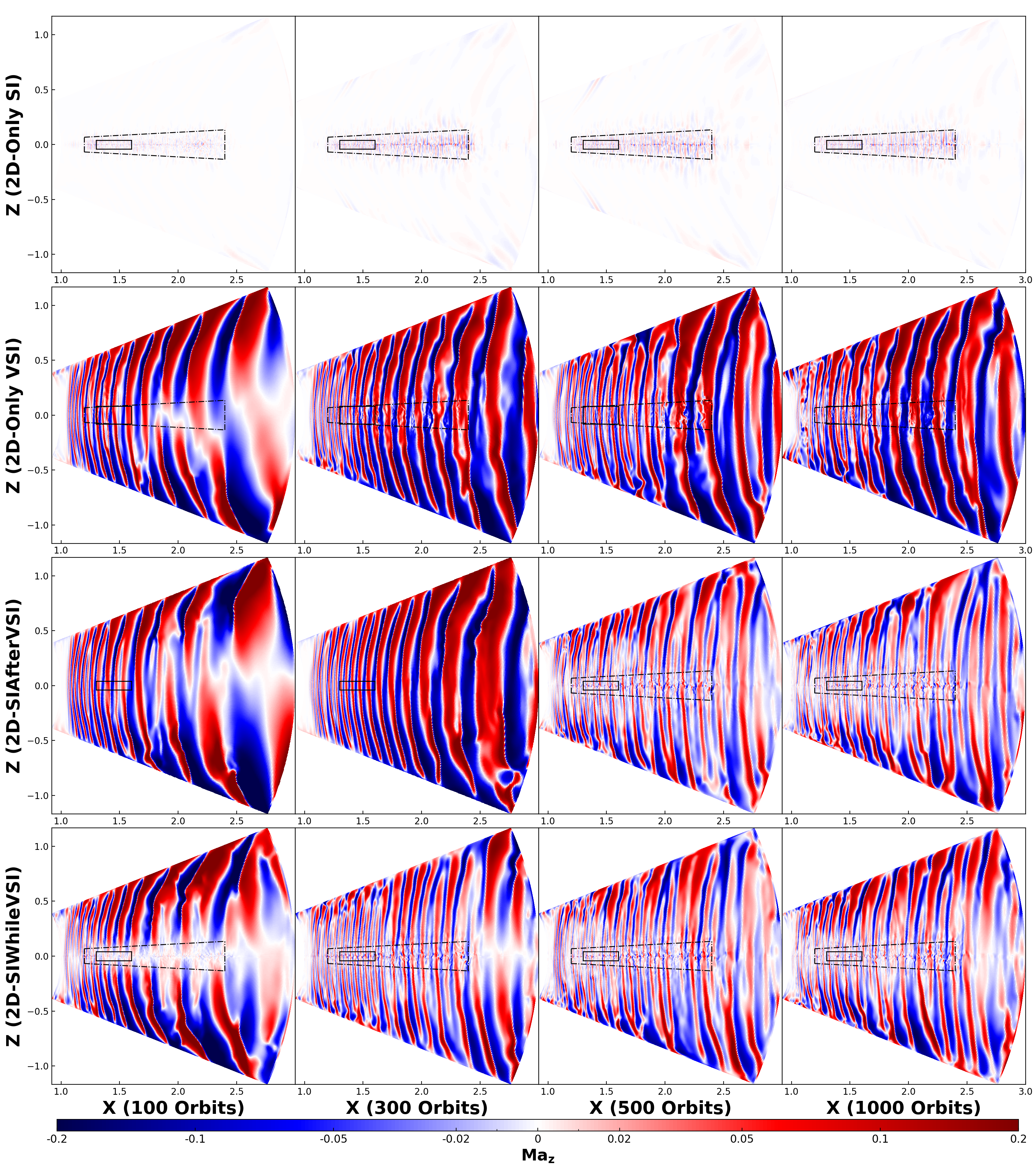}
\caption{2D meridional snapshots of the gas vertical Mach number $Ma_z$ for all 2D models at time $t=100$, $300$, $500$, and $1000$ orbits (from left to right). From top to bottom, the panels correspond to the following runs: ``2D-Only SI'', ``2D-Only VSI'', ``2D-SIAfterVSI'', and ``2D-SIWhileVSI''. The black solid rectangles highlight the zoomed-in region shown in Figure~\ref{fig:2D_dust_ratio}, while the black dashed-dotted rectangles indicate the area with the finest grid.}
\label{fig:2D_velocity_z}
\end{figure*}

\begin{figure*}[htp]
\centering
\includegraphics[scale=0.75]{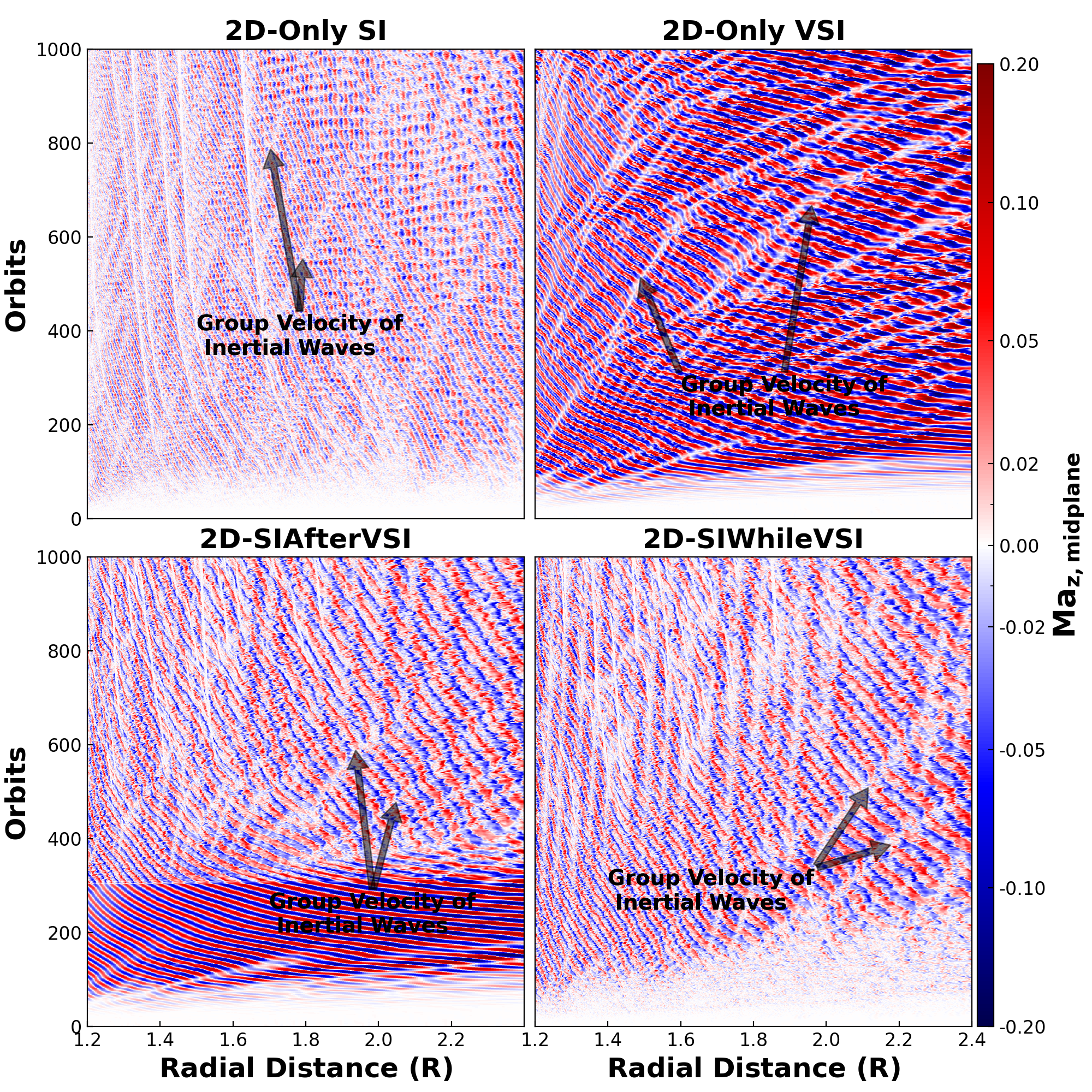}
\caption{The space-time ($R$-$t$) evolution of the vertical Mach number $Ma_z$ at the midplane for various 2D models. The $x$-axis represents the radial distance $R$, while the $y$-axis corresponds to time, measured in orbital periods at the reference radius $R_0$. The white tracks in the vertical Mach number plot result from the transition of two adjacent zones of inertial waves traveling at the group velocity with black arrows highlighting their outward propagation.
}
\label{fig:2D_space_time_vel_z}
\end{figure*}

\begin{figure*}[htp]
\centering
\includegraphics[scale=0.45]{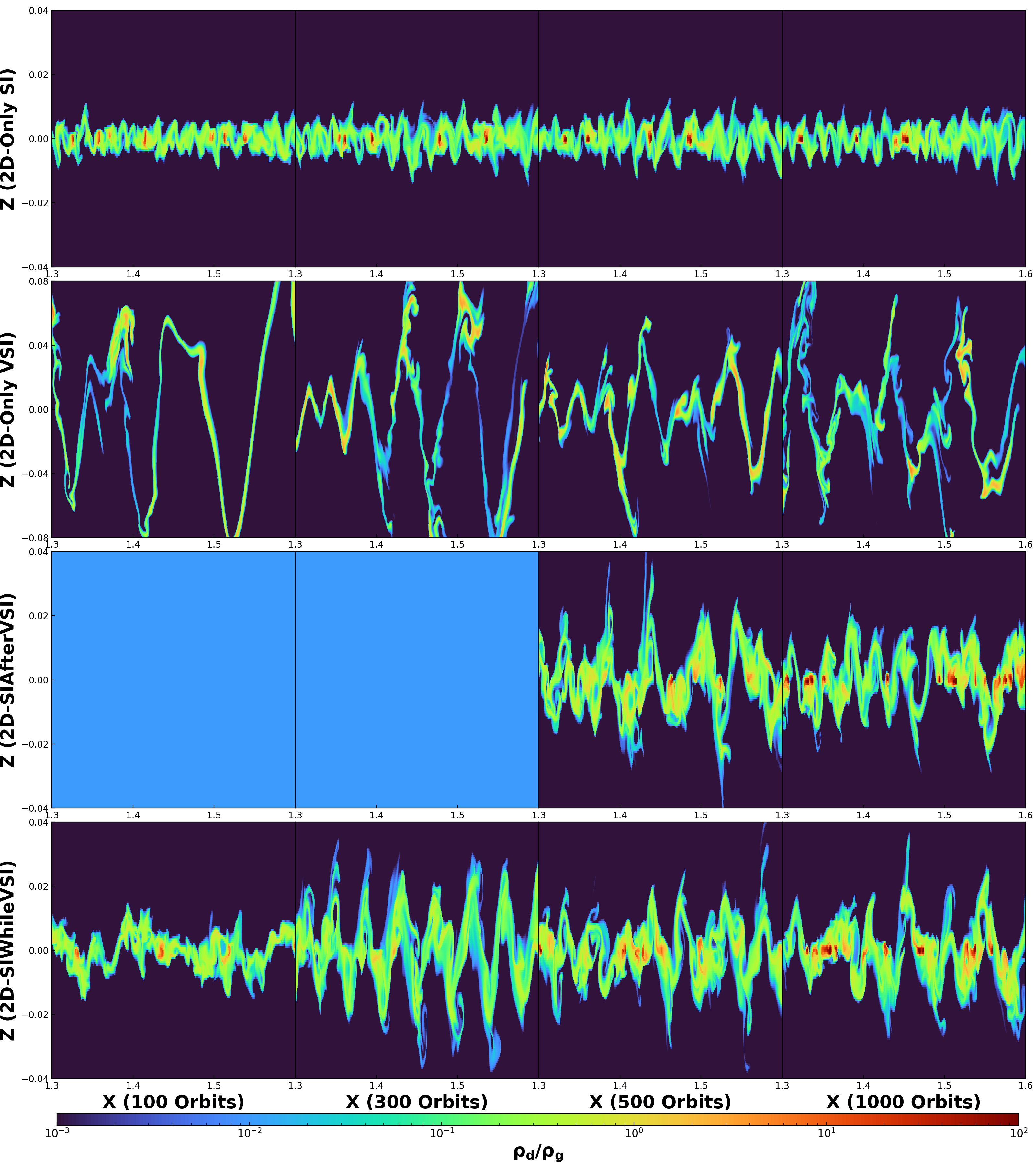}
\caption{Zoomed-in snapshots of the dust-gas density ratios $\rho_\text{d}/\rho_\text{g}$ in the central midplane regions of the simulation domain at time $t=100, 300, 500$ and $1000$ innermost orbits (same as those in Figure~\ref{fig:2D_velocity_z} but in regions enclosed in the black solid rectangles).}
\label{fig:2D_dust_ratio}
\end{figure*}

\begin{figure*}[htp]
\centering
\includegraphics[scale=0.70]{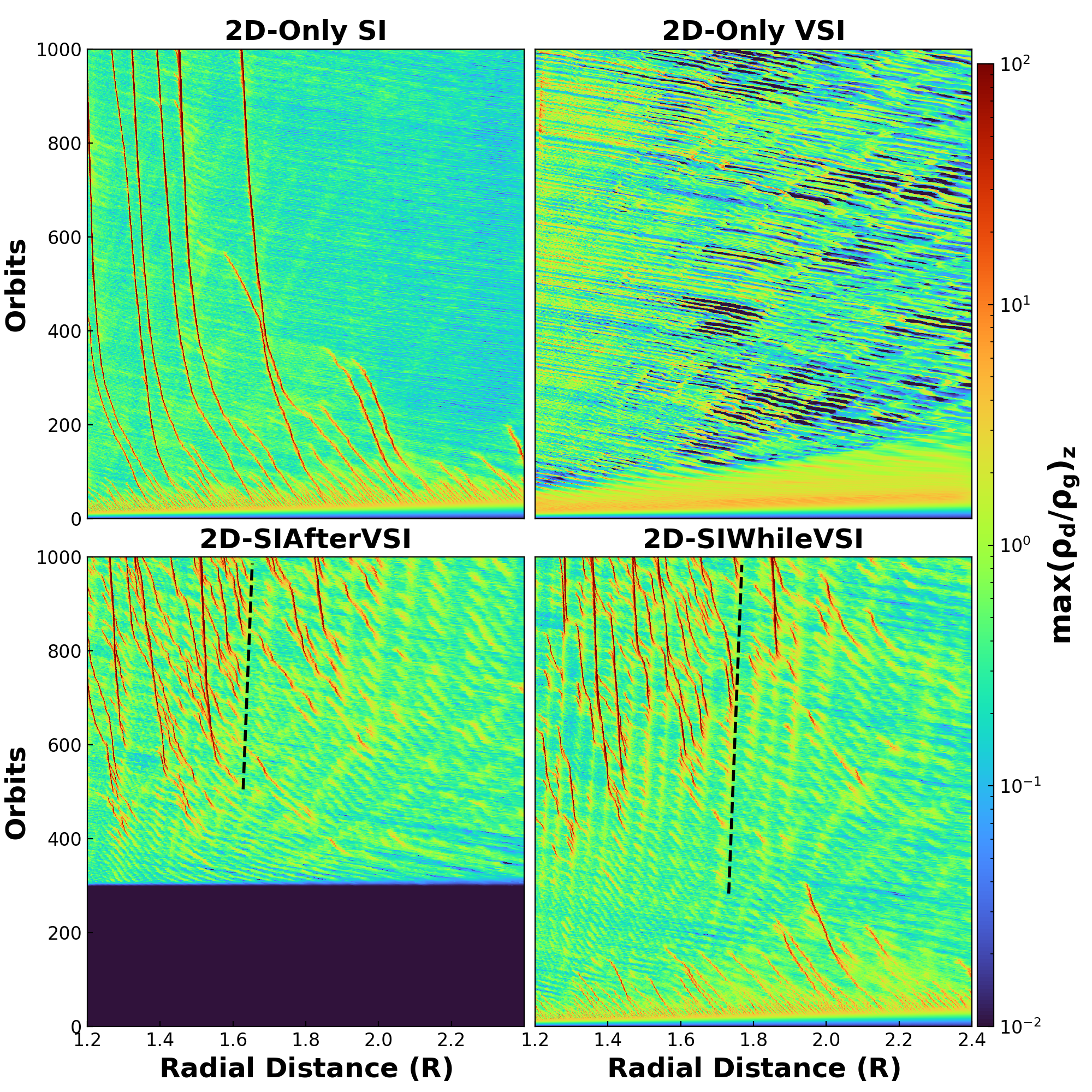}
\caption{Similar to Figure~\ref{fig:2D_space_time_vel_z}, this figure illustrates the space-time evolution of the maxima of dust-gas density ratios, $\max{(\rho_\text{d}/\rho_\text{g})_z}$, for the 2D models. The $x$-axis represents the radial distance, $R$, while the $y$-axis indicates time in orbital periods at the reference radius, $R_0$. Black dashed lines mark regions associated with weak zonal flows and fluctuations in the radial pressure gradient (see Figure~\ref{fig:2D_space_time_pressure}).}
\label{fig:2D_space_time_dust_ratio}
\end{figure*}

\begin{figure*}[htp]
\centering
\includegraphics[scale=0.70]{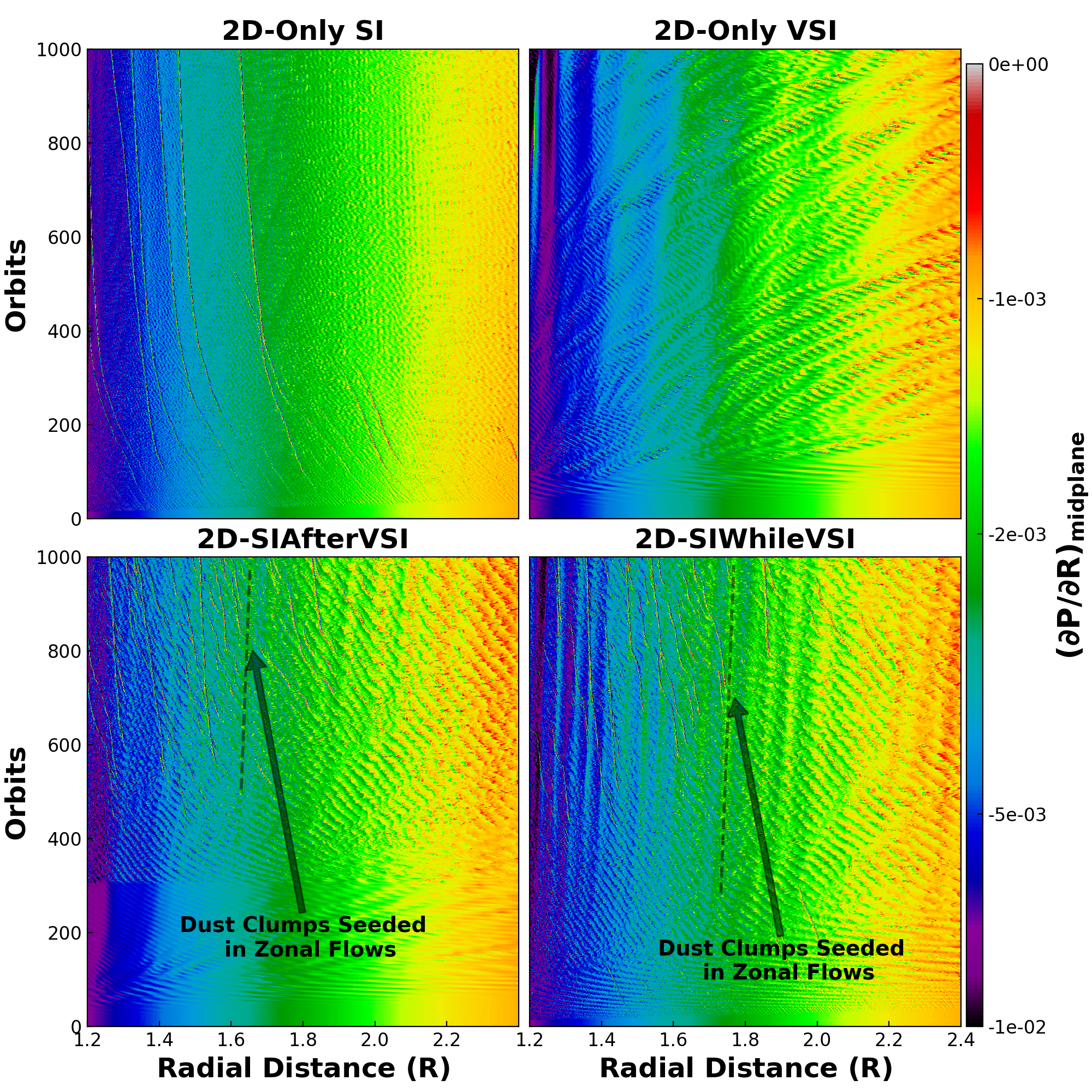}
\caption{Similar to Figure~\ref{fig:2D_space_time_vel_z}, but illustrating the space-time evolution of the radial pressure gradient at the midplane, $(\partial P / \partial R)_\text{midplane}$, for the 2D models. As in Figure~\ref{fig:2D_space_time_dust_ratio}, the black dashed lines highlight regions where weak zonal flows in the radial pressure gradient emerge.
Dust clumps are seeded in these weak zonal flows (pressure gradient fluctuations) in ``2D-SIAfterVSI'' and ``2D-SIWhileVSI'', but note that the pressure gradient remains negative in these regions.}
\label{fig:2D_space_time_pressure}
\end{figure*}

\begin{figure*}[htp]
\centering
\includegraphics[scale=0.58]{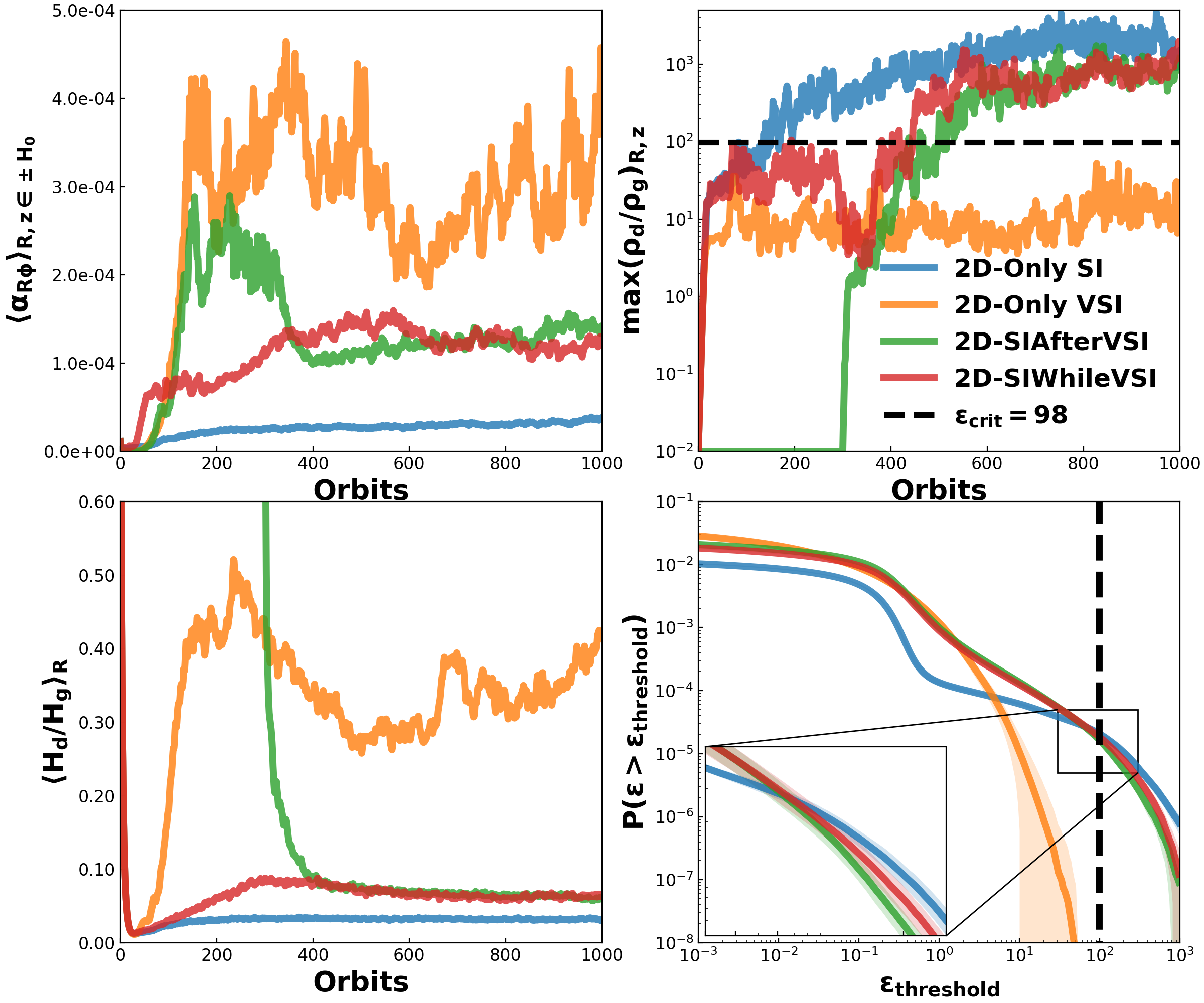}
\caption{Top-Left: The temporal evolution of the radially and vertically averaged  ($z\in \pm H_0$, where $H_0 = 0.08$) Reynolds stress, $\langle \alpha_{R\phi} \rangle_{R, z \in \pm H_0}$, calculated using Equations~\ref{eq:alpha_Rphi}. Top-Right: The temporal evolutions of the maxima of dust-gas density ratios. Bottom-Left: The temporal evolution of the radially averaged dust scale height $\langle  H_\text{d}/H_\text{g} \rangle_R$ (Equation~\ref{eq:averaging_Hd}) for 2D models. Bottom-Right: The cumulative distribution functions (CDFs) of dust-gas density ratios counted by the number of cells for 2D models. The solid lines and the shaded regions represent the mean value and the temporal standard deviations calculated from 800 to 1000 orbits. The inset panel provides a magnified view of the CDFs in the range $\epsilon = 30$ to $\epsilon = 300$. The black dashed lines indicated the critical ratios for planetesimals formation (Equation~\ref{eq:crit_ratio}).}
\label{fig:2D_lines}
\end{figure*}

\subsection{General Turbulence Properties}~\label{subsec:2DTurbulence}

We begin by discussing the general properties of SI and VSI turbulence. As the most prominent feature of VSI turbulence is large-scale vertical motion, Figure~\ref{fig:2D_velocity_z} presents time snapshots of the gas vertical Mach number for various models, while Figure~\ref{fig:2D_space_time_vel_z} shows the space-time ($R$-$t$) plots of the gas vertical Mach number. Additionally, the top-left panel of Figure~\ref{fig:2D_lines} illustrates the temporal evolution of the radially and vertically averaged Reynolds stress, $\alpha_{R\phi}$.

It is evident that the ``2D-Only SI'' model is distinct from the models with VSI. SI with $St = 0.1$ typically grows at a rate of approximately $0.1 \Omega_\text{K}^{-1}$ to $0.4 \Omega_\text{K}^{-1}$, depending on the dust abundance~\citep{YoudinJohansen2007,JohansenYoudin2007}. In the absence of VSI, the vertical Mach number remains small, around $Ma_z \lesssim 0.01$, as shown in Figures~\ref{fig:2D_velocity_z} and~\ref{fig:2D_space_time_vel_z}. The turbulence generated by SI alone is weak, with Reynolds stress being $\alpha_{R\phi} \sim 3\times 10^{-5}$.

In the ``2D-Only VSI'' model, in the absence of dust feedback, there is only VSI turbulence. VSI turbulence can be categorized into two components: surface modes and body modes~\citep{Nelson2013,BarkerLatter2015}. Surface modes are acoustic waves that occur at high altitudes within the disk and are likely boundary effects. These modes have minimal impact on dust dynamics at the midplane. In contrast, body modes, which are inertial waves propagating within the disk, can significantly levitate dust at the midplane. Among these, body modes with $k_z = 2$ and $k_z = 1$ are referred to as breathing modes and corrugation modes, respectively. Breathing modes create antisymmetric patterns in the vertical Mach number $Ma_z$ ($v_{\text{g},z}\sim0$ at midplane), while corrugation modes produce symmetric patterns in $Ma_z$ with significant vertical motion at the midplane.

The evolution of $Ma_z$ in the ``2D-Only VSI'' model progresses as follows: initially, high-body modes ($k_z > 2$) dominate, followed by breathing modes ($k_z = 2$), as observed at 100 orbits in Figure~\ref{fig:2D_velocity_z}. Finally, corrugation modes ($k_z = 1$) emerge at 300 orbits in Figure~\ref{fig:2D_velocity_z} and dominate afterwards. The vertical wavelength typically corresponds to the vertical extent of disks. The vertical Mach number reaches values of approximately $Ma_z \sim 0.2$. This progression is consistent with previous studies~\citep{Nelson2013,StollKley2014}. Overall, VSI turbulence is strongly anisotropic ($k_R \sim h^{-1} k_z \sim 10 k_z$), resulting in elongated patterns of $Ma_z$. The effective $\alpha_{R\phi}$ lies between $2 \times 10^{-4}$ and $4 \times 10^{-4}$.

The other two runs with VSI+SI, regardless of their initial conditions and simulation procedures, behave similarly once the simulations reach a saturated state. Specifically, during the first 300 orbits of the ``2D-SIAfterVSI'' run, without dust back-reaction, the system develops vigorous gaseous VSI turbulence, similar to the first 300 orbits of ``2D-Only VSI''. After 300 orbits, once mesh refinement is enabled and dust is introduced, the vertical Mach number decreases to $Ma_z \sim 0.05$ due to dust back-reaction. The radial wavenumber $k_R$ of the VSI body modes increases (see Figures~\ref{fig:2D_velocity_z} and~\ref{fig:2D_space_time_vel_z}). This phenomenon can be understood in terms of dusty buoyancy, where dust mass loading increases fluid inertia but not pressure, which serves as a stabilizing effect that resists vertical oscillations~\citep{LinYoudin2017}. As shown in Figure 13 of~\citet{LinYoudin2015}, VSI body modes with larger $k_R$ generally have higher growth rates and longer critical cooling times, making them more resilient to buoyant suppression. Consequently, the presence of dusty buoyancy tends to favor modes with higher $k_R$, leading to a shift in the dominant $k_R$ toward higher wavenumbers.

In ``2D-SIWhileVSI'', the antisymmetric breathing modes persist longer compared to the ``2D-Only VSI'' and ``2D-SIAfterVSI'' models, as seen at 100 orbits in Figure~\ref{fig:2D_velocity_z}. Ultimately, the velocity patterns and the turbulent stress, $\alpha_{R\phi}$, in ``2D-SIWhileVSI'' and ``2D-SIAfterVSI'' become very similar at 500 and 1000 orbits. The $\alpha_{R\phi}$ values induced by VSI+SI are significantly larger than those from SI alone but smaller than those from VSI alone, at approximately $\alpha_{R\phi} \sim 10^{-4}$.

One notable feature in the development of VSI is the generation of inertial waves. In the ``2D-Only VSI'', ``2D-SIAfterVSI'', and ``2D-SIWhileVSI'' models, the body modes of vertical Mach numbers eventually become dominated by corrugation modes. These VSI body modes behave similarly to traveling inertial ``wave trains'', with phase velocity propagating inward and group velocity propagating outward~\citep{BarkerLatter2015,Svanberg2022,Ogilvie2025}. The outward group velocity is most visible when one wave zone with a coherent frequency transitions to the next, which manifests as jumps or discontinuities in the traveling body modes, represented by the white lines extending outward in Figure~\ref{fig:2D_space_time_vel_z} (highlighted by black arrows). Notably, the gas inertial waves are also vigorously present in the SI case, as seen in the space-time plots of vertical Mach numbers for the ``2D-Only SI'' model.

\subsection{Dust Ratios, Scale Heights and Distributions}~\label{subsec:2DDust}

In this subsection, we focus on the properties of dust, as illustrated through the following figures. Figure~\ref{fig:2D_dust_ratio} presents time snapshots of the dust-gas density ratios for various models, while Figure~\ref{fig:2D_space_time_dust_ratio} shows space-time ($R$-$t$) plots of the maximum dust-gas density ratios. The top-right and bottom-left panels of Figure~\ref{fig:2D_lines} display the temporal evolution of the maxima of dust-gas density ratios, $\max{(\rho_\text{d}/\rho_\text{g})_{R,z}}$, and the radially-averaged dust-gas scale height ratios, $\langle H_\text{d}/H_\text{g} \rangle_R$, respectively. The bottom-right panel of Figure~\ref{fig:2D_lines} provides the cumulative distribution functions (CDFs) of the dust-gas density ratios in the 2D simulations. The calculation of the CDFs for the dust distribution using the multifluid method follows the approach outlined in Section 3.5.6 of~\cite{Benitez2019FARGO3D} and Section 3.3.3 of~\cite{HuangBai2022}. The CDFs are computed by counting the number of cells where the dust-gas density ratios exceed specified threshold values. Additionally, Figure~\ref{fig:2D_space_time_pressure} shows the space-time ($R$-$t$) plots of the gas pressure gradient at the midplane.

We begin with the ``2D-Only SI'' model. In the early stages, weak and transient dust clumps form at the midplane (see the 100-orbit snapshots in Figure~\ref{fig:2D_dust_ratio} and the first 100 orbits in Figure~\ref{fig:2D_space_time_dust_ratio}). These clumps either merge or dissipate, eventually evolving into long-lived clumps that can survive for hundreds of orbits. The dust-gas density ratio exceeds the critical threshold of $\epsilon_\text{crit} = 98$ at around 100 orbits (see the top-right panel of Figure~\ref{fig:2D_lines}). The temporal evolution of the dust scale height, shown in the bottom-left panel of Figure~\ref{fig:2D_lines}, reveals that the mean dust scale height is approximately 3\% of the gas scale height ($\langle H_\text{d}/H_\text{g} \rangle_R \sim 0.03$), while in regions that form dust clumps, it can decrease to as low as 1\% ($H_\text{d,clump} \sim 0.01 H_\text{g}$).

Next, the ``2D-Only VSI'' model generates vigorous VSI turbulence dominated by vertical motion, and the dust layer in this model has a much larger scale height compared to other models ($\langle H_\text{d}/H_\text{g} \rangle_R \sim 0.3$ to $0.4$, see also Figure~\ref{fig:2D_dust_ratio}). 
The maxima of the dust-gas density ratios never exceeds the critical threshold $\epsilon_\text{crit} = 98$. On the other hand, the VSI turbulence generates zonal flows (see the pressure fluctuations in Figure~\ref{fig:2D_space_time_pressure}) that slow down (but not completely halt) the radial drift of dust and enhance the dust-gas density ratios.

The ``2D-SIAfterVSI'' model allows the SI to develop within the existing VSI turbulence. This indeed occurs, albeit at a later time after introducing particles and dust feedback. Compared to the ``2D-Only SI'' model, the ``2D-SIAfterVSI'' model produces more dust clumps (Figure~\ref{fig:2D_space_time_dust_ratio}), though only a fraction of these clumps exceed the critical dust-gas density ratio $\epsilon_\text{crit}$. These clumps tend to form in the weak zonal flows driven by VSI turbulence (see the black dashed lines in Figures~\ref{fig:2D_space_time_dust_ratio} and~\ref{fig:2D_space_time_pressure}), consistent with the findings of~\cite{SchaferJohansen2020},~\cite{SchaferJohansen2022} and~\cite{Schafer2025}. Note that the zonal flows exhibit as radial pressure variations that do not form pressure maxima. The dust scale height in the model ``2D-SIAfterVSI'' is $\langle H_\text{d}/H_\text{g} \rangle_R \sim 0.1$, in between the ``2D-Only SI'' and ``2D-Only VSI'' models (see discussion next).

In the ``2D-SIWhileVSI'' model, there is competition between the initial growth of the SI and the development of VSI. With $St = 0.1$, the growth rate of the SI is higher than that of the VSI corrugation modes~\citep[typically $|q|h\Omega_K^{-1} \sim 0.05 \Omega_K^{-1}$;][]{BarkerLatter2015}. As a result, dust clumps form at the midplane due to SI during the first 100 orbits of the ``2D-SIWhileVSI'', similar to the ``2D-Only SI'' model. However, as VSI corrugation modes develop at the midplane, these dust clumps are lifted up and eventually disrupted by VSI turbulence. Due to dust feedback, the dust scale height in this model, although higher than in the pure SI case, is still lower by a factor of $\gtrsim 2$ compared to the case without dust feedback~\citep{Lin2019}. This reduction is caused by an extra buoyancy force induced by dust loading, which effectively weakens the VSI. After a new balance is reached between dust settling and stirring from modified VSI turbulence, the SI and associated dust clumping re-emerge. These dust clumps are also formed in the pressure gradient fluctuations of zonal flows in VSI (see the black dashed lines in Figure~\ref{fig:2D_space_time_pressure}). As seen in Figures~\ref{fig:2D_dust_ratio},~\ref{fig:2D_space_time_dust_ratio}, and~\ref{fig:2D_lines}, the morphology of the maxima in dust-gas density ratios and dust-gas scale height ratios closely resembles that of the ``2D-SIAfterVSI'' model.

Finally, we examine the properties of dust clumping in more detail by analyzing the cumulative distribution functions (CDFs) of these models, shown in the bottom-right panel of Figure~\ref{fig:2D_lines}. In the ``2D-Only VSI'' model, the CDF never reaches $\epsilon_\text{crit}$. In the ``2D-Only SI'', ``2D-SIAfterVSI'', and ``2D-SIWhileVSI'' models, about $10^{-5}$ to $10^{-4}$ of the cells exceed the critical density, with the highest dust concentration exceeding $10^3$, potentially triggering planetesimal formation. Among them, the CDFs of dust distribution for the ``2D-SIWhileVSI'' and ``2D-SIAfterVSI'' models are nearly identical.
In contrast, the high-density part of the CDF in the ``2D-Only SI'' model is flatter with a heavier tail at $\epsilon_\text{threshold} > 60$, implying that the pure SI clumps are likely more massive, potentially leading to
more top-heavy planetesimal initial mass function.

\subsection{Dust Released Initially or Later?}~\label{subsec:VSIandSI2D}

The interplay between the SI and VSI has recently been investigated by~\cite{SchaferJohansen2020}, ~\cite{SchaferJohansen2022} and~\cite{Schafer2025} using 2D simulations. They suggested that the outcome depends on timing of dust release. When dust is released from the beginning (``SIWhileVSI''), turbulence at the midplane tends to be dominated by SI, which grows faster. In contrast, if dust is released after the VSI has saturated (``SIAfterVSI''), VSI turbulence prevails at the midplane. They also noted that dust clumping is stronger in the latter case.

However, our findings, shown in Figures~\ref{fig:2D_velocity_z} to~\ref{fig:2D_lines}, suggest that the final outcomes of the ``SIWhileVSI'' and ``SIAfterVSI'' models are essentially identical. This discrepancy is likely due to the different sizes of the meridional domain used in the simulations:~\cite{SchaferJohansen2020, SchaferJohansen2022} and~\cite{Schafer2025} employed a domain size of $L_z = 2 H_\text{g} \sim 4 H_\text{g}$, while our study uses a larger domain of $L_z = 7 H_\text{g} \sim 10 H_\text{g}$. Since the free energy driving the VSI originates from vertical shear, and vertical shear increases with height from the midplane~\citep{BarkerLatter2015}, it is conceivable that vertical domain size needs to be sufficiently large ($\Delta z \gtrsim 6 H_\text{g}$) to properly capture the non-linear properties of the VSI turbulence (Appendix~\ref{app:different_H}, see also ~\cite{Fukuhara2023}). Even though the growth rates of VSI body modes is smaller than those of SI, the turbulence at the midplane is still controlled by VSI.

In reality, the onset of the SI and VSI depends on dust properties via aerodynamic coupling and opacity. While we anticipate that the ``SIAfterVSI'' scenario is more plausible as it likely takes more time for dust to reach relatively large Stokes number ($St = 0.1$), employed. The situation is simplified thanks to our finding. Therefore, in the 3D simulations presented in this and the subsequent papers, we simply adopt the ``SIWhileVSI'' approach to release dust at the beginning of the simulations.

\section{3D Global Simulations Results}~\label{subsec:3DRunsVSIandRWI}

In this section, we present the results from two 3D simulations to investigate the 3D VSI turbulent properties and how the dust feedback modify the gas and dust behaviors.

\begin{figure*}[htp]
\centering
\includegraphics[scale=0.50]{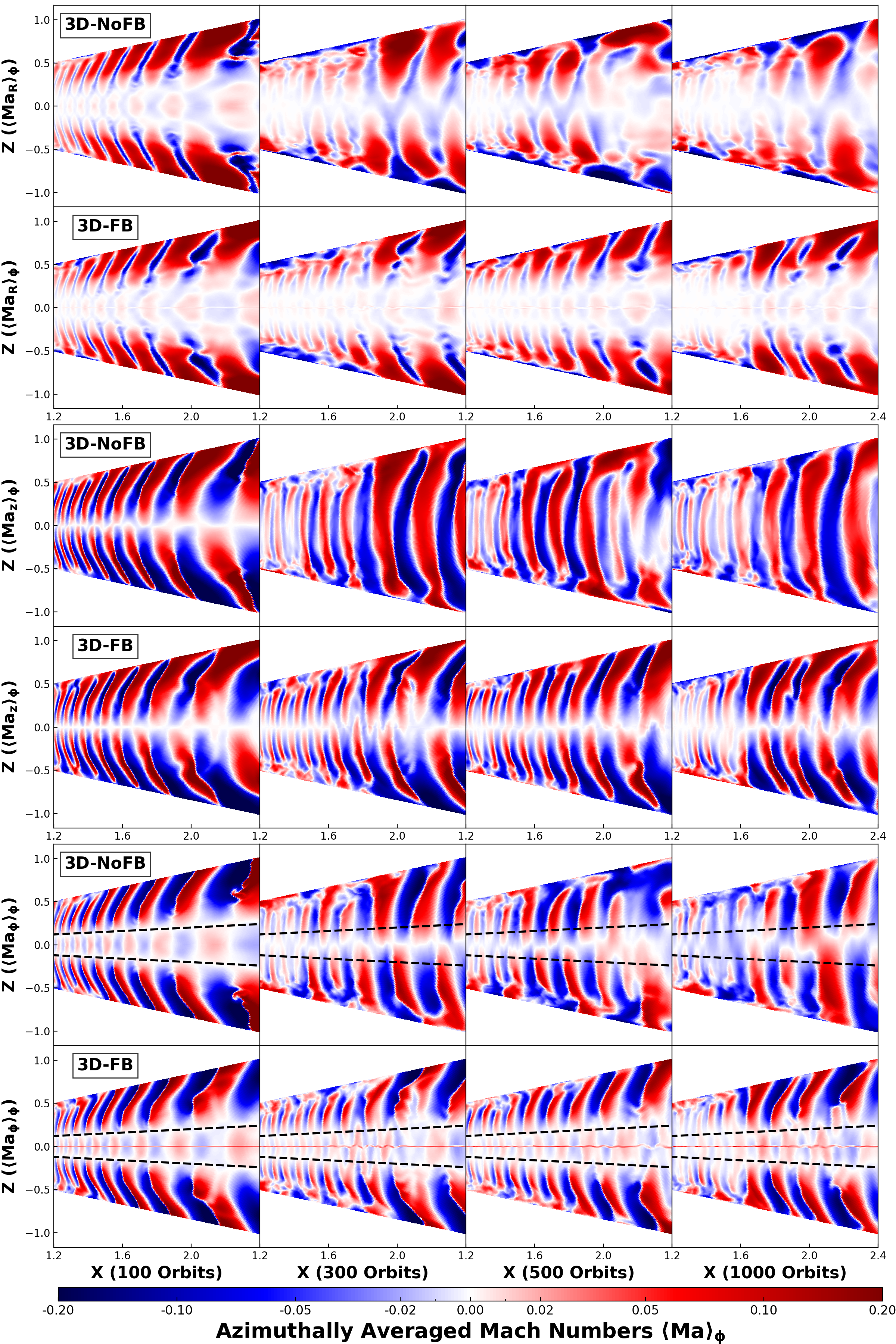}
\caption{The azimuthally averaged Mach number $\langle Ma\rangle_\phi$ (Equation~\ref{eq:Ma}) along three directions are depicted for 3D models at various time intervals. From left to right, the panels represent $100$, $300$, $500$, and $1000$ orbits (at $R_0$). The top two rows display $\langle Ma_R\rangle_\phi$ for ``3D-NoFB'' and ``3D-FB'', the middle two rows depict $\langle Ma_z\rangle_\phi$ for ``3D-NoFB'' and ``3D-FB'', while the bottom two rows illustrate $\langle Ma_\phi\rangle_\phi$ for ``3D-NoFB'' and ``3D-FB''. The zonal flows generated by VSI (red and blue regions around the midplane in $\langle Ma_\phi\rangle_\phi$) are marked between black dashed lines.}
\label{fig:3D_Mach_numbers}
\end{figure*}

\begin{figure*}[htp]
\centering
\includegraphics[scale=0.60]{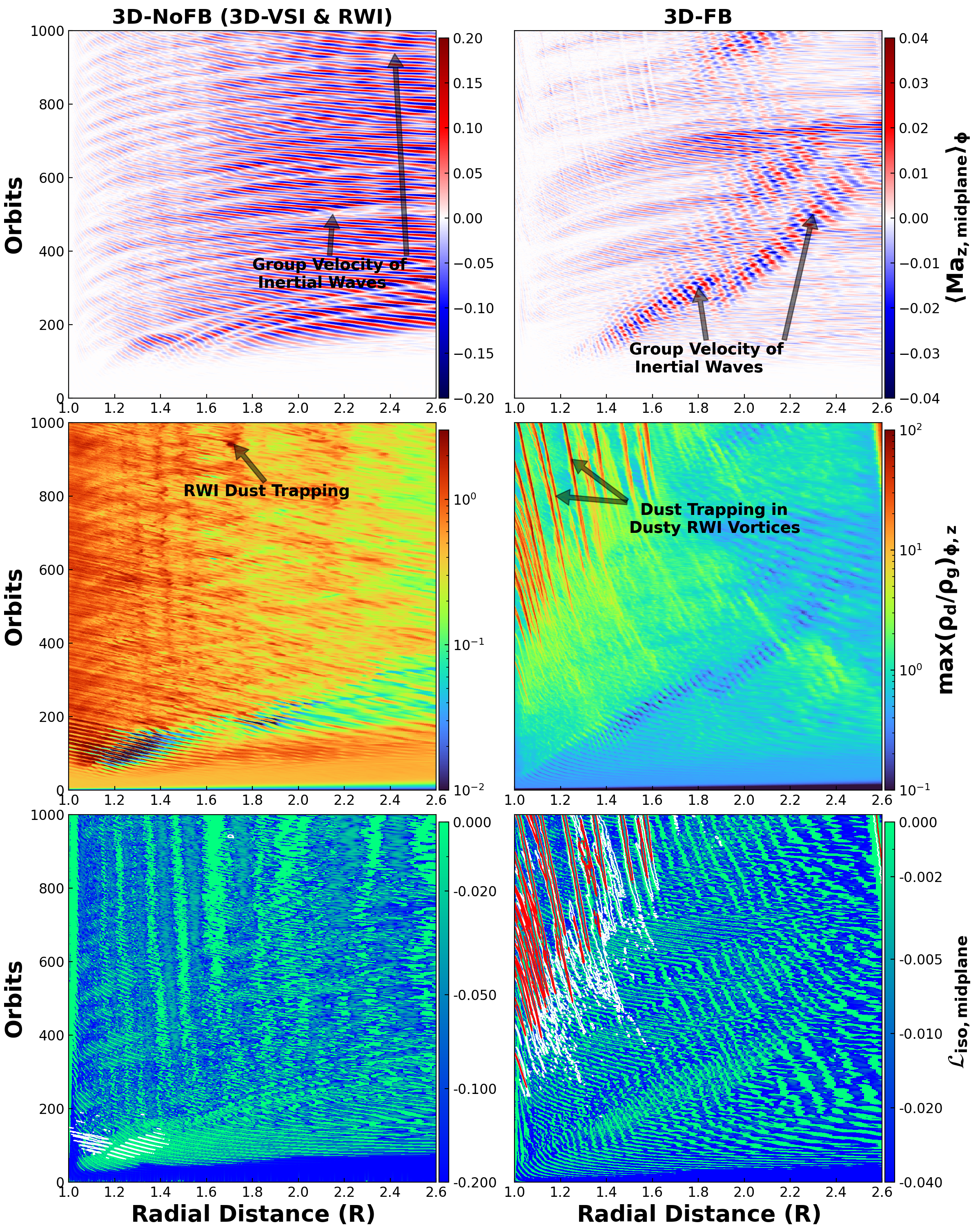}
\caption{The space-time ($R$-$t$) plots of the azimuthally averaged vertical Mach number at the midplane, $\langle Ma_{z,\text{midplane}} \rangle_\phi$ (top row), the maxima of dust-gas density ratios, $\max{(\rho_\text{d}/\rho_\text{g})_{\phi,z}}$ (middle row), and the locally isothermal RWI Key function, $\mathscr{L}_\text{iso,midplane}$ (Equation~\ref{eq:Lfunc_iso}, bottom row), for the ``3D-NoFB'' and ``3D-FB'' models. In the bottom row, the white and red contours indicate regions where the dust-gas density ratios are equal to 3 and 10, respectively.}
\label{fig:3D_space_time_general}
\end{figure*}

\begin{figure*}[htp]
\centering
\includegraphics[scale=0.55]{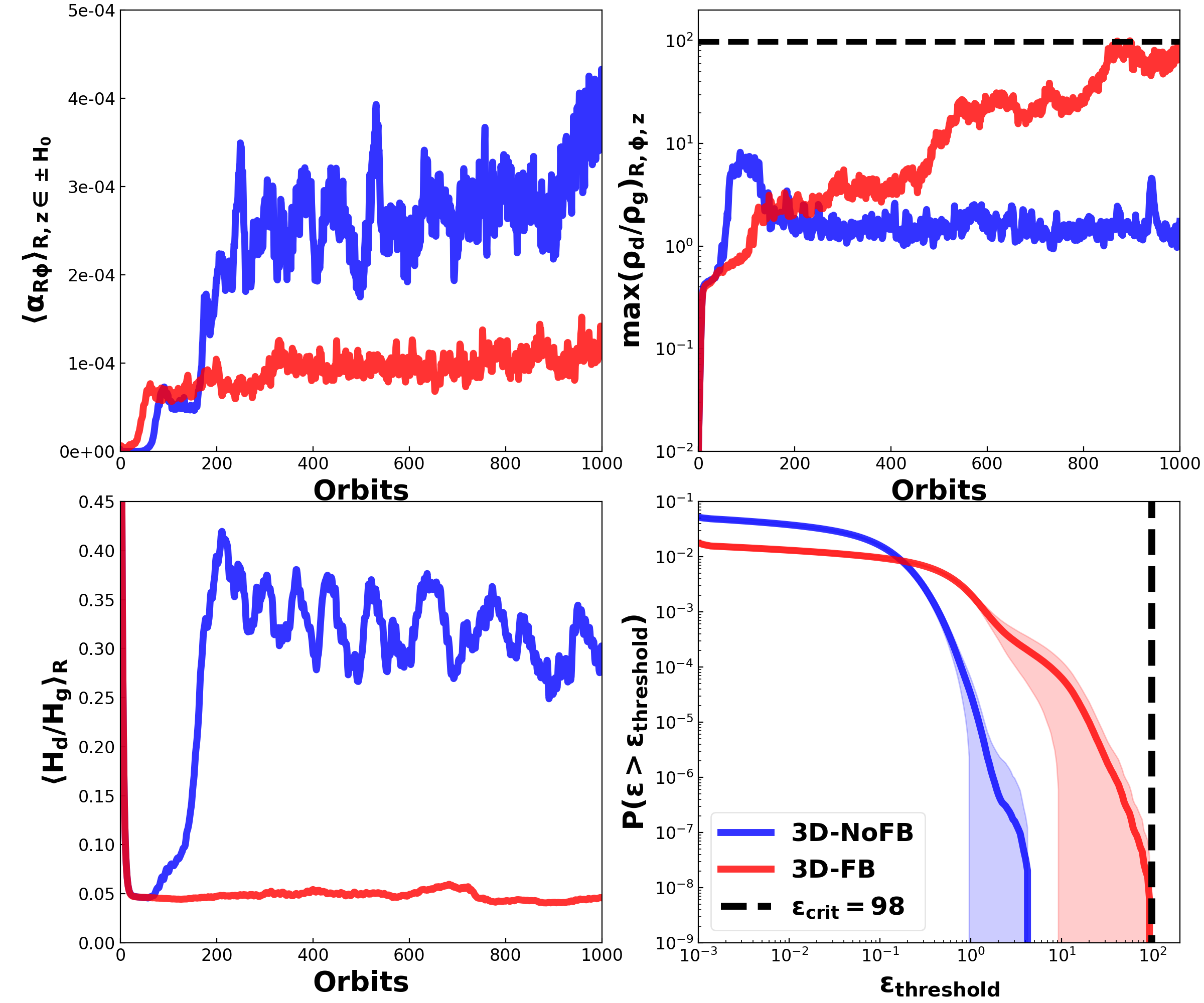}
\caption{Similar to Figure~\ref{fig:2D_lines}, but for the 3D runs (``3D-NoFB'' and ``3D-FB'').}
\label{fig:3D_lines}
\end{figure*}

\begin{figure*}[htp]
\centering
\includegraphics[scale=0.60]{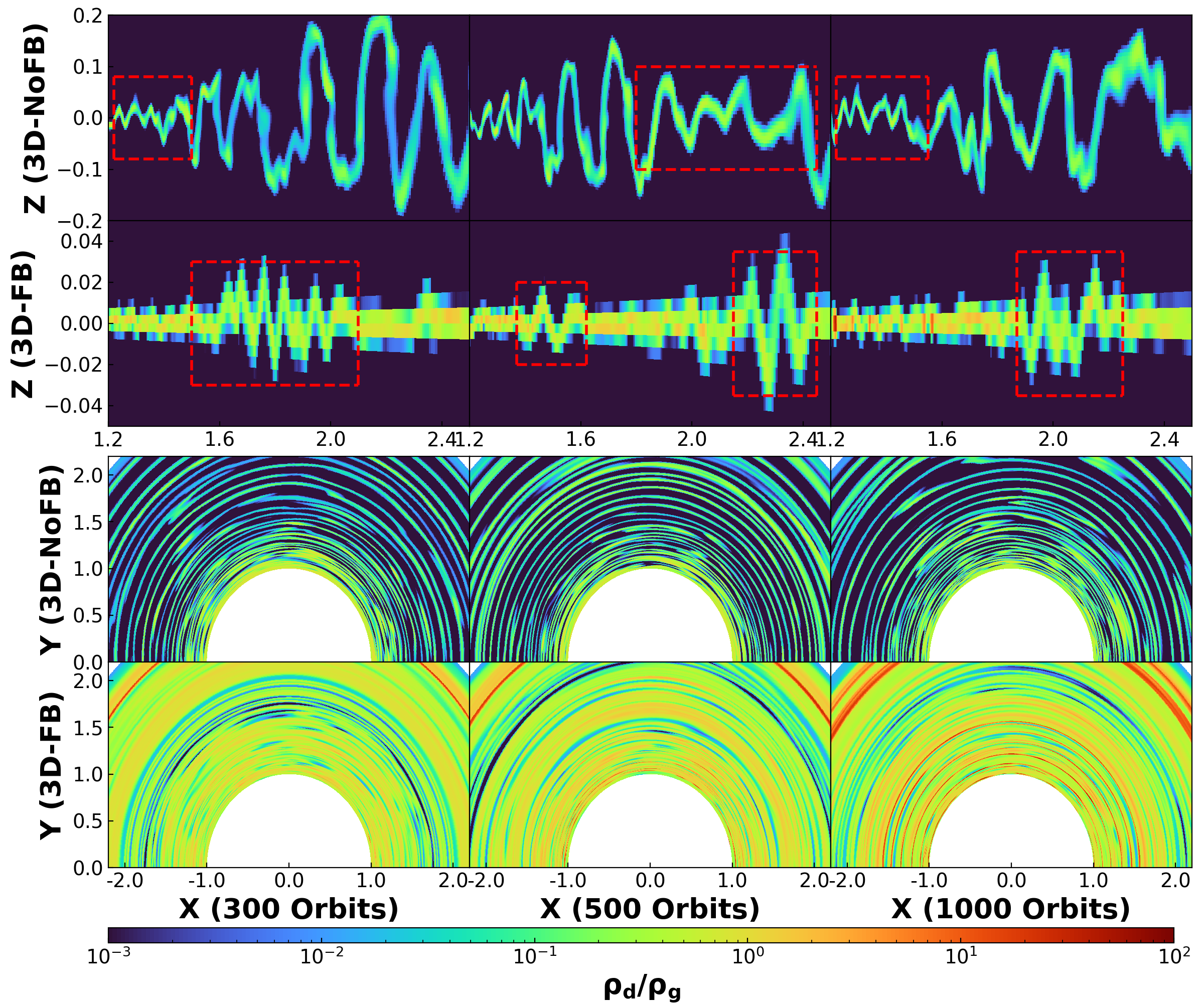}
\caption{The dust-gas density ratios $\rho_\text{d}/\rho_\text{g}$ on the $X$-$Z$ section ($\phi = \pi/2$, $Y = 0$, top six panels) and the $X$-$Y$ section ($\theta = \pi/2$, $Z = 0$, midplane, bottom six panels) for the ``3D-NoFB'' and ``3D-FB'' models at various times are shown. The choice of the azimuthal angle is arbitrary. From the left to the right columns, the panels correspond to 300, 500 and 1000 orbits. The first and third rows correspond to the ``3D-NoFB'' model, while the second and fourth rows correspond to the ``3D-FB'' model. The red dashed rectangles highlight regions where the dust layer is located at the boundary between two adjacent zones of inertial waves which propagates at the their group velocity. In the ``3D-NoFB'' model
where corrugation modes dominate, this results in less dust stirring, whereas in the ``3D-FB'' model where breathing modes dominate, this leads to more pronounced dust stirring.}
\label{fig:3D_dust_ratio}
\end{figure*}

\begin{figure}[htp]
\centering
\includegraphics[scale=0.35]{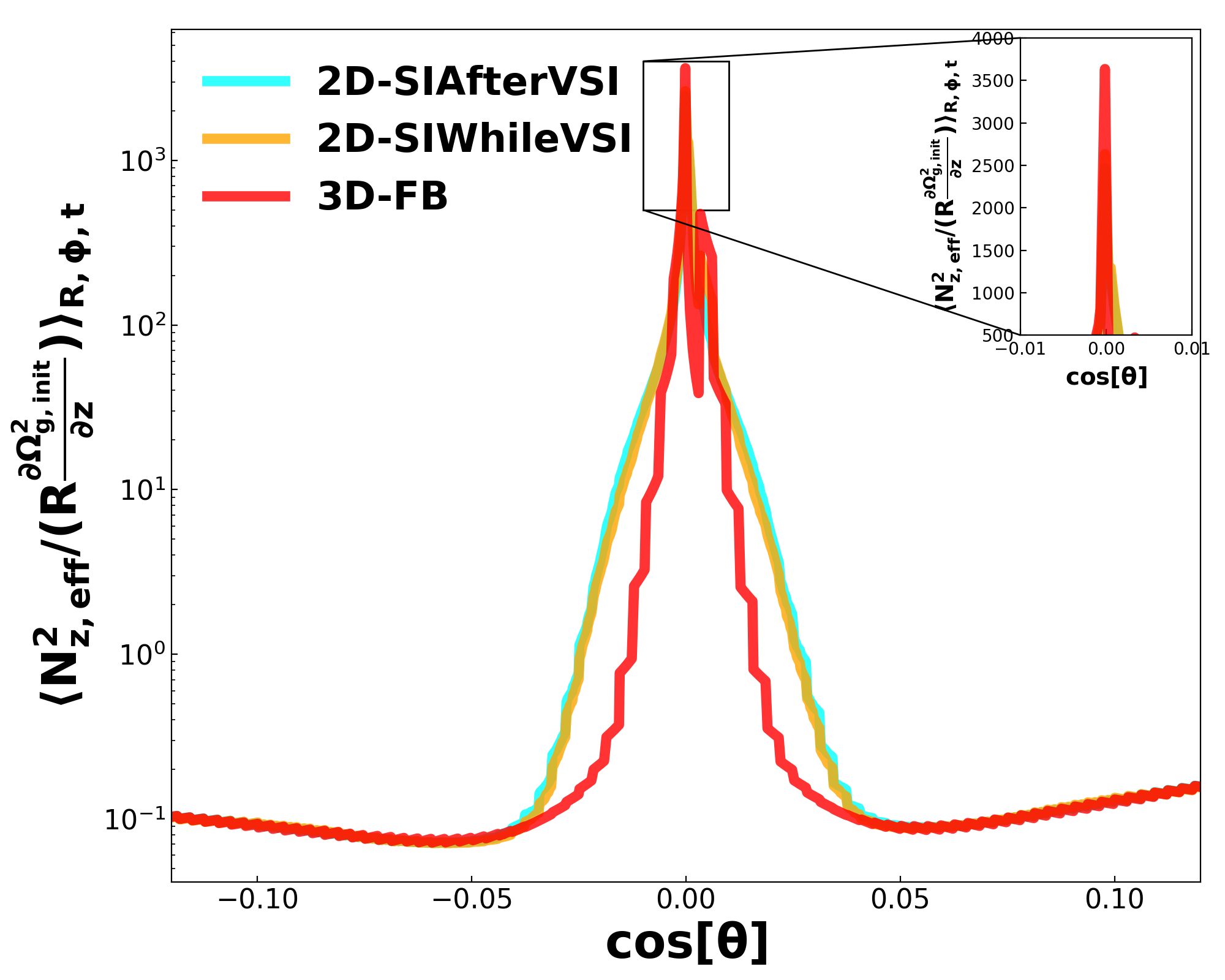}
\caption{The radially, azimuthally, and temporally averaged ratio of the squared dusty buoyancy frequency, $N_\text{z,eff}^2$ (Equation~\ref{eq:buoyancy}), to the temperature-induced vertical shear, $R \partial \Omega^2_\text{init} / \partial z$, for the ``2D-SIAfterVSI'', ``2D-SIWhileVSI'', and ``3D-FB'' models. The inset panel provides a zoomed-in view around the midplane ($\theta = \pi/2$).
}
\label{fig:3D_dusty_buoyancy}
\end{figure}

\begin{figure*}[htp]
\centering
\includegraphics[scale=0.60]{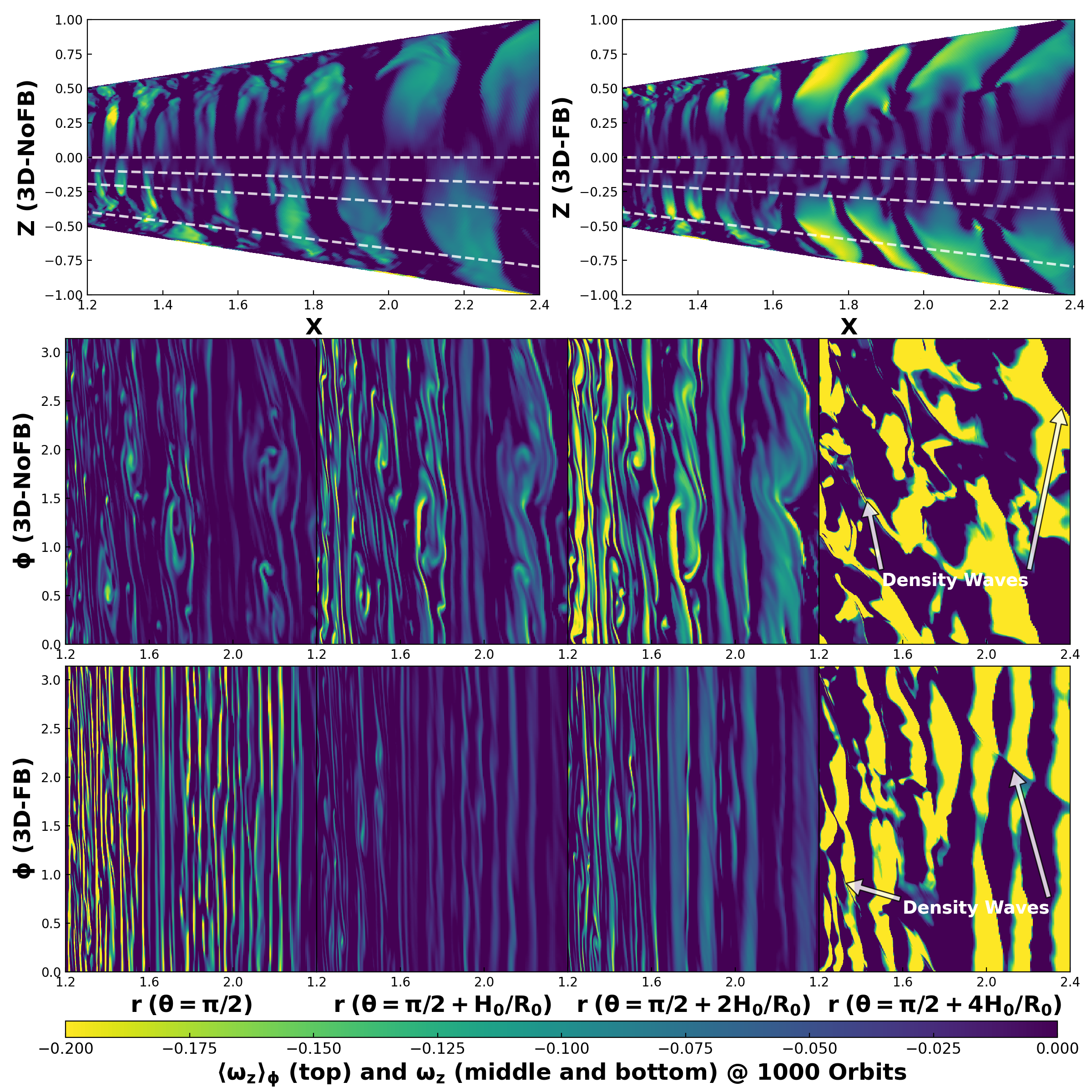}
\caption{The top two panels show the azimuthally averaged vertical vorticity, $\langle \omega_z \rangle_\phi$ (Equation~\ref{eq:vorticity}), at 1000 orbits for the ``3D-NoFB'' (left) and ``3D-FB'' (right) models. The middle and bottom four panels present various cuts (indicated by white dashed lines in the top two panels) of the vertical vorticity $\omega_z$ at 1000 orbits. Given the symmetry or antisymmetry of the vertical vorticity about the midplane, we specifically select the following meridional angles: $\theta = \pi/2$ ($z = 0$, midplane), $\pi/2 + H_0 / R_0$, $\pi/2 + 2H_0 / R_0$, and $\pi/2 + 4H_0 / R_0$, where $H_0 / R_0 = 0.08$. Density waves at $\theta = \pi/2 + 4H_0 / R_0$ are indicated by white arrows.}
\label{fig:3D_vorticity_z}
\end{figure*}

\subsection{General Turbulence Properties}~\label{subsec:3DTurbulence}

We begin by discussing the general properties of VSI turbulence. Figure~\ref{fig:3D_Mach_numbers} displays snapshots of the azimuthally averaged gas Mach numbers. An azimuthal slice of the gas Mach number is also presented in Appendix~\ref{app:axisymmetry}, indicating that VSI turbulence is predominantly axisymmetric. The top two panels of Figure~\ref{fig:3D_space_time_general} show space-time ($R$-$t$) plots of the azimuthally averaged vertical Mach number at the midplane, $\langle Ma_{z,\text{midplane}} \rangle_\phi$. Furthermore, the top-left panel of Figure~\ref{fig:3D_lines} illustrates the temporal evolution of the radially and azimuthally averaged Reynolds stress, $\langle \alpha_{R\phi} \rangle_{R,z}$.

In the ``3D-NoFB'' model, without dust feedback, this model recapitulates earlier works of~\citet{StollKley2016,Flock2017b,MangerKlahr2018,Manger2020} and~\cite{Flock2020}.
After azimuthal averaging, the Mach number patterns display outward-bending modes that counteract vertical shear~\citep{BarkerLatter2015}, similar to the 2D case. At altitudes beyond $|z|\sim3H_\text{g}$, the Mach number in all components reaches $\sim 0.2$, likely due to surface modes induced by boundary effects. Initially dominated by high-order body modes, the pattern of vertical Mach number then evolves into breathing modes and eventually into corrugation modes. At the midplane, the vertical Mach number is significantly larger than the radial Mach number, with $\langle Ma_{z,\text{midplane}}\rangle_\phi/\langle Ma_{R,\text{midplane}}\rangle_\phi \sim |q|h^{-1} \sim 5$~\citep{Nelson2013,StollKley2014}. Zonal flows are visible at the midplane in the azimuthal Mach number (see the bottom-left panels in Figures~\ref{fig:3D_Mach_numbers}), while the radial Mach number alternates between accretion and decretion flows at the midplane. The Reynolds stress $\alpha_{R\phi}$ is approximately $10^{-4}$ around the midplane, and increases to about $5\times10^{-4}$ when integrated over the entire disk. All these features are consistent with previous works.

In the early stages of the ``3D-FB'' model (within 100 orbits), the Mach numbers and turbulent $\alpha_{R\phi}$ are similar to those in the ``3D-NoFB'' model. In both cases, the vertical Mach numbers are primarily dominated by breathing modes. However, as time progresses, breathing modes persist in the ``3D-FB'' model, but corrugation modes do not appear—this is a key difference from the ``3D-NoFB'' model. This difference will be explained in Section~\ref{subsec:buoyancy}. Dust feedback accelerates the gas azimuthal velocities, as seen in the bottom panels of Figure~\ref{fig:3D_Mach_numbers} (visible as red lines at the midplane). Unlike the ``3D-NoFB'' model, the gas at the midplane systematically move outward in the ``3D-FB'' model, which is entirely due to dust feedback as gas gains angular momentum from the dust. Interestingly, the integrated Reynolds stress is significantly lower than that in the ``3D-NoFB'' model by a factor of $\sim4$, which exemplifies that dust generally inhibits gas motion and hence reduce turbulence.

\subsection{Dust Ratios, Scale Heights and Distributions}~\label{subsec:3DDust}

To investigate the dust properties, we present in Figure~\ref{fig:3D_dust_ratio} snapshots of the dust-gas density ratios at a specific azimuthal slice ($\phi = \pi/2$) and at the midplane ($z = 0$). The middle two panels of Figure~\ref{fig:3D_space_time_general} show space-time ($R-t$) plots of the maximum dust-gas density ratios. Additionally, the top-right, bottom-left, and bottom-right panels of Figure~\ref{fig:3D_lines} illustrate the temporal evolution of the maxima of the dust-gas density ratios, the radially averaged dust-gas scale height ratios, and the cumulative distribution functions (CDFs) of the dust, respectively.

By 100 orbits in the ``3D-NoFB'' model, the VSI begins to generate zonal flows that divide the dust layer into several rings with weak dust concentrations (Figure~\ref{fig:3D_dust_ratio}). When VSI turbulence fully saturates at about 300 orbits, corrugation modes lift the dust layer to higher altitudes, with dust scale heights reaching $\langle H_\text{d}/H_\text{g}\rangle_R = 0.3 \sim 0.4$, slightly smaller than that of ``2D-Only VSI''.
As the VSI body modes exhibit as a train of inertial waves propagating inwards (similar to the 2D cases), the vertical gas motion de-correlate at the location where one wave zone transition to the next (which moves at the group velocity of inertial waves),
leading to a reduced dust scale height (highlighted by the red rectangle in Figure~\ref{fig:3D_dust_ratio}). The maximum dust-gas density ratio reaches only approximately unity. While vortices are present (Figure~\ref{fig:3D_vorticity_z}, and see later discussion), they only lead to weakly non-axisymmetric dusty rings (bottom three panels in Figure~\ref{fig:3D_dust_ratio}), and fail to efficiently trap dust with $St = 0.1$.

In the ``3D-FB'' model, the dust profile shown in Figure~\ref{fig:3D_dust_ratio} indicates that the dust layer is much thinner compared to the ``3D-NoFB'' model (note the different scale used in the $y-$axis). The smaller dust scale heights can also be seen in the radially-averaged dust-gas scale height ratios in Figure~\ref{fig:3D_lines}, indicating $H_{\text{d}} \sim 0.05 H_{\text{g}}$. This value is even lower than $H_{\text{d}} \sim 0.08 H_{\text{g}}$ observed in 2D models of VSI turbulence with dust feedback (Figures~\ref{fig:2D_lines}). This further reduced dust scale layer thickness is associated with the suppression of corrugation mode and the dominance of the breathing mode, where the anti-symmetric vertical profile of $v_z$ reduces gas vertical motion. As a result, even without further mesh refinement, large dust-gas density ratios are observed at the midplane ($\max{(\rho_\text{d}/\rho_\text{g})} = 10 \sim 100$). The dusty rings at the midplane in the ``3D-FB'' model are also more axisymmetric compared to those in the ``3D-NoFB'' model. We thus anticipate the SI to be able to operate should we increase the resolution.

From the temporal evolution of the maximum dust-gas density ratios and the CDFs in Figure~\ref{fig:3D_lines}, we see that the dust density largely remain low in the ``3D-NoFB'' model. The initial rapid increase of the dust-gas density ratio within the first 100 orbits results from dust settling before the VSI corrugation modes are fully developed, and this is followed by dust spreading as the VSI turbulence is saturated. In the ``3D-FB'' model, with dust feedback, we again find that the dust-gas density ratio is substantially enhanced, reaching the critical threshold even though SI is not expected to operate at root-level resolution without mesh refinement. In the rest of the discussion, we will conduct more detailed analysis on how dust back-reaction alters the properties of the VSI and RWI turbulence, while a more comprehensive study of dust concentration and clumping with mesh refinement will be deferred to a follow up paper.

\subsection{Dusty Buoyancy and VSI Body Modes}~\label{subsec:buoyancy}

As stated earlier, a main distinction between the ``3D-NoFB'' and ``3D-FB'' models is the dominance of breathing modes in the latter while the corrugation modes are suppressed due to dust back-reaction. To look into this further, the role of dust mass loading could be characterized by its effective buoyancy frequency squared $N_\text{z,eff}^2$ (\ref{eq:buoyancy}). Its effectiveness could be countered by temperature induced vertical shear which can be characterized by $R \partial \Omega^2_\text{g,init}/\partial z$~\citep{Lin2019}.
Figure~\ref{fig:3D_dusty_buoyancy} presents the radially, azimuthally and temporally averaged ratio between the effective dusty buoyancy and the temperature-induced vertical shear in the ``2D-SIAfterVSI'', ``2D-SIWhileVSI'' and ``3D-FB'' models. The temperature-induced vertical shear increases with the altitude of the disks, while the dusty buoyancy is mostly concentrated at the midplane. Clearly, in both 2D and 3D runs, dusty buoyancy dominates over shear in the midplane layer. Compared to 2D cases, the dusty effective buoyancy of ``3D-FB'' are more concentrated around the midplane.

Another way to look into the suppression of corrugation mode is to note that body modes with odd-symmetry ($v_z\neq0$ at midplane, including corrugation but not breathing modes) require communication between the top and bottom halves of the disk~\cite[see Figure 10 in ][]{BarkerLatter2015}. The strong dusty buoyancy weakens or inhibits this communication, and hence the corrugation mode. As the fundamental body modes (breathing and corrugation) contain most of the kinetic energy~\citep{Nelson2013,BarkerLatter2015,Umurhan2016},
only breathing modes are observed in the ``3D-FB'' run. We also note that when additional effects such as magnetic fields or stellar irradiation are considered, the corrugation modes may also be suppressed, leaving the breathing modes dominant~\citep{CuiBai2020,ZhangZhu2024}.

The strength of turbulence stirring, at least in the vertical direction, is known to be weaker in 3D VSI compared to 2D VSI turbulence~\citep{MangerKlahr2018}~\citep[see also Figure 3 in][]{StollKley2014}. As a result, dust tends to settle more in 3D than in 2D, and this also holds when dust feedback is included, as clearly seen in the dust-gas scale ratio panels in Figure~\ref{fig:2D_lines} and~\ref{fig:3D_lines}. As a result, dusty buoyancy in 3D cases is stronger and more concentrated in the midplane layer (Figure~\ref{fig:3D_dusty_buoyancy}), and the late stages of 2D VSI turbulence are still dominated by weak corrugation modes (Figure~\ref{fig:2D_velocity_z}).

\subsection{Vertical Vorticity and RWI Vortices}

The most prominent feature in 3D VSI turbulence is the presence of numerous vorticies. Such vortices are anti-cyclonic with pressure maxima at the center, and hence provide a favorable environment for trapping dust, facilitating dust concentration and growth. The vortices in VSI turbulence are generated by the RWI as a secondary instability resulting from VSI-induced zonal flows~\citep{Richard2016,MangerKlahr2018,Manger2020}.
We visualize the vortices by showing the azimuthally-averaged vertical vorticity $\langle \omega_z \rangle_\phi$ and the horizontal distribution of vertical vorticity at different heights at 1000 orbits of run ``3D-NoFB'' and ``3D-FB'' in Figure~\ref{fig:3D_vorticity_z}.
Note that as the gas is non-barotropic, $\omega_z$ is not conserved, and hence inevitably exhibit vertical structures.

In model ``3D-NoFB'', the midplane region is characterized by modest-sized vortices, with typical vertical vorticity around $\omega_{z,\text{midplane}} \sim -0.1$. By comparing with the bottom-left panel of Figure~\ref{fig:3D_space_time_general}, we can identify that the locations where vortices are found are clearly associated with the maximum of the RWI key function $\mathscr{L}_{\text{iso,midplane}}$ (Equations~\ref{eq:Lfunc_iso}), confirming their origin as RWI vortices resulting from the zonal flows. The vortices also slowly migrate inward during the simulation. Despite of very weak dust clumping, there is clear association of dust trapping pattern with the pattern in $\mathscr{L}_{\text{iso,midplane}}$ (and hence the zonal flows) when comparing with the middle left panel of  Figure~\ref{fig:3D_space_time_general}. From  Figure~\ref{fig:3D_vorticity_z}, we further see that such vortices extend to $2 H_\text{g} \sim 3 H_\text{g}$, where $\omega_z$ increases with height with its value roughly doubles at $2 H_\text{g} \sim 3 H_\text{g}$. The pattern of $\omega_z$ follows the patterns of $v_{g,R}$ and $v_{g,\phi}$ and are anti-symmetric about the midplane. At $\theta = \pi/2 + 4H_0/R_0$, the surface modes of VSI create a strong turbulent environment with high vertical vorticity, where spiral density waves emitted by vortices~\citep{Bodo2005,Paardekooper2010} could be observed.
All these results are consistent with previous works~\citep{Richard2016,StollKley2016,MangerKlahr2018,Manger2020}.

More vorticity rings at the midplane ($\theta = \pi/2$) are observed in the ``3D-FB'' model. These rings are much more closely packed (typical separation $\sim 0.3 H_\text{g}$), exhibiting stronger vertical vorticity ($\omega_z \sim -0.5$) and are less asymmetric compared to those of ``3D-NoFB''. Higher up ($\theta\gtrsim\pi/2 + 2H_0/R_0$), the vertical vorticity remains similar to that in the ``3D-NoFB'' model.  When comparing with the space-time plots of $\mathscr{L}_{\text{iso,midplane}}$ (bottom-right panel in Figure~\ref{fig:3D_space_time_general}) and the dust-gas density ratios (middle-right panel in Figure~\ref{fig:3D_space_time_general}) of the ``3D-FB'' run, we observe that the local maxima of $\mathcal{L}_\text{iso,midplane}$ align with the location of dusty ring edges (see the white and red contours in the bottom-right panel of Figure~\ref{fig:3D_space_time_general}). Dust feedback leads to gas reaching Keplerian or even super-Keplerian speeds within dust clumps, while gas outside these clumps remains sub-Keplerian. The strong velocity variations activate the dusty RWI at the edges of the dusty rings~\citep{LiuBai2023}. Unlike the classic gaseous RWI, the dusty RWI tends to generate multiple smaller, isolated vortices rather than forming a single large vortex~\citep{FuLi2014, PierensLin2019, HuangLiIsella2020, HsiehLin2020, YangZhu2020, ChanPaardekooper2024}. Dusty RWI also tends to preserve the ring-like structures compared to the gaseous RWI vortices~\citep{LiuBai2023}, as we observe from our simulation.

\subsection{Stability of 3D Vortices and Energy Cascade}~\label{subsec:vortices}

\begin{figure*}[htp]
\centering
\includegraphics[scale=0.35]{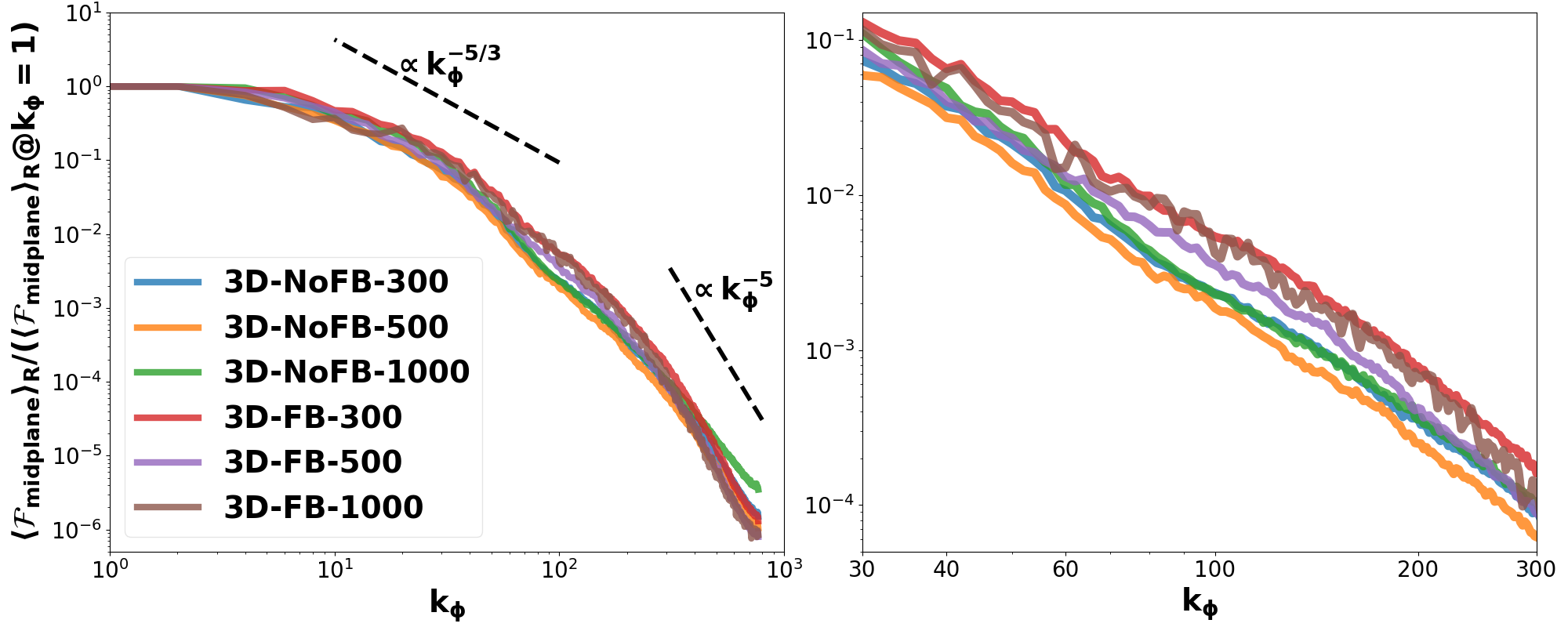}
\caption{The left panel displays the radially averaged ($R \in [1.2, 2.4]$) power spectrum $\langle \mathcal{F} \rangle_R$ (Equation~\ref{eq:fft_kinerg}), normalized by its value at $k_\phi = 1$, for the azimuthal wavenumber ($k_\phi$) of the total kinetic energy density $\delta E_\text{kin}$ (Equation~\ref{eq:kinerg}) at the midplane in two 3D models (``3D-NoFB'' and ``3D-FB''). In the ``3D-NoFB'' model, since dust is only an imperfect tracer, $\delta E_\text{kin}$ is computed solely from the gas contribution. The right panel presents a zoomed-in view focusing on azimuthal wavenumbers $k_\phi$ between 30 and 300. Different colors represent different snapshots for ``3D-NoFB'' and ``3D-FB''.}
\label{fig:3D_fft_kinetic_energy}
\end{figure*}

Gaseous vortices in 3D are known to be susceptible to the Elliptic Instability (EI), which is a result of the resonance between the vortex motion and inertial waves. This could generate turbulence within the vortices and lead to their destruction~\citep{LesurPapaloizou2009}, especially for Vortices with small aspect ratios (Equation~\ref{eq:aspect}, $\chi \lesssim 4$).

In our ``3D-NoFB'' model, the aspect ratios of vortices range from 2 to 12, and all vortices exhibit vertical structures, as shown in Figure~\ref{fig:3D_vorticity_z}. Smaller vortices tend to decay within several orbits, likely due to the EI. By contrast, larger vortices can develop internal turbulent cores while persist for hundreds of orbits until the end of the simulations. This longevity may be attributed to the insufficient resolution in our simulations to fully capture the EI within these larger vortices, aligning with findings from previous 3D simulations of VSI turbulence~\citep{Richard2016,MangerKlahr2018,Manger2020}.

The dynamics in the ``3D-FB'' model are more intricate. The distribution of aspect ratios of vortices in the ``3D-FB'' model is similar to that in the ``3D-NoFB'' model at high altitudes, although the body modes of ``3D-FB'' are primarily dominated by breathing modes. The thin dust layer divides the disk into top and bottom halves, inhibiting communication between these regions. Breathing modes in the ``3D-FB'' model generate anticyclonic bands, indicating that the vortices retain vertical structures (see Figure~\ref{fig:3D_vorticity_z}). Previous 2D studies have shown that dust feedback can disrupt vortex structures~\citep{FuLi2014,Raettig2015}. However,~\cite{Lyra2018} found that dust feedback primarily affects vortex structures at the midplane, while columnar structures can persist at higher altitudes. In the ``3D-FB'' model, no large-scale vortex structures are observed at the midplane; instead, several narrow, weakly asymmetric vorticity rings induced by dusty RWI are present.

The high filling fraction of vortices of various sizes, combined with the potential for internal turbulence within them, motivates a closer examination of the kinetic energy spectrum in the azimuthal ($\phi$) direction. The radially averaged Fourier power spectra of the total kinetic energy density (Equation~\ref{eq:kinerg}) at various times ($t =$ 300, 500, and 1000 orbits) at the midplane are shown in Figure~\ref{fig:3D_fft_kinetic_energy}. They all reveal a double cascade: a shallow power law (\(\propto k_\phi^{-5/3}\)) at low wavenumbers and a steep power law (\(\propto k_\phi^{-5}\)) at high wavenumbers. These results are consistent with those presented in Figure~13 of~\cite{MangerKlahr2018} and Figure~6 of~\cite{Manger2020}, suggestive of an upward Kolmogorov cascade and a downward enstrophy cascade, as discussed there.
In the ``3D-FB'' model, dust feedback further enhances the axisymmetry of VSI turbulence (see Figure~\ref{fig:3D_vorticity_z}), shifting the inverse Kolmogorov cascade regime to higher wavenumbers ($k_\phi \simeq 30 \sim 300$) compared to the ``3D-NoFB'' model (see the right panel of Figure~\ref{fig:3D_fft_kinetic_energy}).

\section{Conclusion and Discussion}~\label{sec:summary}

In this work, we use the multifluid dust module in Athena++ and conduct a set of 2D and 3D simulations to investigate the dust-gas interaction under VSI turbulence in PPDs. Our 2D simulations employ SMR which allows us to simultaneously resolve the SI, together with sufficient $\theta-$domain to reduce the influence of vertical boundaries. We focus on the interplay between the VSI and the SI, and the results serve as a benchmark for understanding more realistic results in 3D. Our 3D simulations are restricted to root-level resolution. While not able to resolve the SI, it allows us to study the how dust back-reaction affects the properties of the VSI turbulence for the first time in 3D, and sets the stage for our followup work which will resolve the SI as well.

Based on the 2D simulations, we can draw the following conclusions:
\begin{itemize}
    \item Vertical Shear Instability (VSI) and Streaming Instability (SI) can coexist within protoplanetary disks (Figure~\ref{fig:2D_velocity_z} and Figure~\ref{fig:2D_dust_ratio}). SI could survive within VSI turbulence, produces significant dust overdensities that reach the Roche density, facilitating planetesimal formation (Figure~\ref{fig:2D_space_time_dust_ratio}). These overdensities are seeded in regions with weaker pressure gradient fluctuations generated by VSI-induced zonal flows (Figure~\ref{fig:2D_space_time_pressure}).
    \item VSI turbulence is highly anisotropic ($k_R \gg k_z$) and primarily dominated by corrugation modes. When dust feedback is included, dust settles into a thinner layer, leading to an increase in radial wavenumbers ($k_R$ becomes larger) and a weakening of the corrugation modes (see Figure~\ref{fig:2D_velocity_z} and Figure~\ref{fig:2D_space_time_vel_z}).
    \item Using a larger vertical domain ($\Delta z \gtrsim 6 H_\text{g}$), the timing of dust release does not affect the final simulation outcomes (Figure~\ref{fig:2D_lines}), contrary to the findings of~\cite{SchaferJohansen2020},~\cite{SchaferJohansen2022} and~\cite{Schafer2025}. Although the growth rates of SI exceed those of VSI body modes at the midplane, turbulence remains dominated by VSI (see Figure~\ref{fig:2D_velocity_z} and Figure~\ref{fig:2D_dust_ratio}).
\end{itemize}

For 3D simulations, we find that:
\begin{itemize}
    \item In the 3D run without dust feedback (``3D-NoFB''), VSI turbulence at the midplane initially features high-order body modes, which later transition to being dominated by breathing modes and eventually corrugation modes. By contrast, dust feedback in ``3D-FB'' suppresses corrugation modes, leaving breathing modes as the dominant turbulence feature (Figure~\ref{fig:3D_Mach_numbers}).
    \item Vertical stirring is weaker in 3D VSI, leading to thinner dust layer compared to the 2D counterparts (with or without dust feedback). This contributes to the suppression of VSI corrugation modes due to enhanced dust mass loading/buoyancy.
    \item The VSI turbulence generates weak zonal flows (Figure~\ref{fig:3D_Mach_numbers}) that become unstable to the RWI (see Figure~\ref{fig:3D_space_time_general}), leading to the formation of multiple vortices. Dust back-reaction leads to the formation of multiple vorticity rings that exhibit higher vorticity and more compact spacing compared to the dust-free case. These rings further concentrate dust and generate weak and small dusty vorticies at the midplane dust layer. On the other hand, the columnar vortex structures are preserved beyond the dust layer (Figure~\ref{fig:3D_vorticity_z}).

\end{itemize}

In our 3D simulations, we do not have sufficient resolution to fully resolve SI at the midplane in the ``3D-FB'' model. In a following work, we will conduct a 3D global simulation with mesh refinement to showcase the coexistence of VSI, SI and RWI, and investigate the dust clumping under 3D self-consistent turbulence in PPDs.

Our work represents a steady step forward toward understanding the dust dynamics under more realistic gas dynamics in the outer PPDs. With the current focus being the VSI turbulence, we have made several assumptions to simplify the setup. First, we use a simple locally isothermal equation of state. In reality, disk temperature is largely governed by stellar irradiation, featuring a heated surface layer and cooler disk midplane with different effective cooling time across the column. Radiation transport is needed to properly capture more realistic thermodynamics, which could significantly alter the properties of the VSI~\citep{ZhangZhu2024}. Second, we do not include magnetic field, and it is known that the VSI turbulence can co-exist with the MHD wind~\citep{CuiBai2020} and modest level of the magnetorotational instability (MRI) driven turbulence~\citep{CuiBai2022}. In addition, we only consider one single dust species with a constant Stokes number, without accounting for a dust size distribution~\citep{Bai2010dynamics,Krapp2019,YangZhu2021SI} and dust evolution~\citep{LiLi2020Coagulation,TominagaTanaka2023,Pfeil2023}. Incorporating these factors in future work will allow for a more comprehensive understanding of dust dynamics in PPDs.

\nolinenumbers
\section*{Acknowledgments}
We thank Orkan Umurhan, Andrew Youdin, Hui Li, Eugene Chiang, Min-Kai Lin, Chao-Chin Yang, Shengtai Li, Can Cui, Rixin Li and Shangjia Zhang for helpful discussions. This work is supported by the National Science Foundation of China under grant No. 12233004, 12325304, 12342501, and the China Manned Space Project with NO. CMS-CSST-2021-B09. P. Huang acknowledges the Canadian Grant NSERC ALLRP 577027-22.

\vspace{5mm}

\software{astropy~\citep{Robitaille2013astropy}, Athena++~\citep{Stone2020,HuangBai2022}}

\appendix

\section[]{Symbols}\label{app:symbols}

Table~\ref{tab:symbols} lists the symbols used in this study.

\begin{table}[]
\centering
\caption{Parameters List}
\begin{footnotesize}
\begin{tabular}{ccc}
\toprule
\hline
Symbol          & Definition                                    & Description                                                   \\
\hline
$r,\;\theta,\;\phi$                                 &                                                                                          & Spherical Polar Coordinate                                   \\
$X,\;Y,\;Z$                                         & $X\equiv r\sin \theta \cos \phi,\;Y\equiv r\sin \theta \sin\phi,\;Z\equiv r\cos \theta $ & Cartesian Coordinate                                         \\
$R,\;\phi,\;z$                                      & $R\equiv r \sin \theta,\; z\equiv r \cos \theta$                                         & Cylindrical Coordinate                                       \\
$k_R,\;k_\phi,\;k_z$                                &                                                                                          & Wavenumbers in Cylindrical Coordinate                        \\
$\rho_\text{g},\;\;\rho_\text{d}$                   &                                                                                          & Densities of Gas and Dust                                    \\
$\boldsymbol{v}_\text{g},\;\boldsymbol{v}_\text{d}$ &                                                                                          & Velocities of Gas and Dust                                   \\
$\Omega_\text{g},\;\Omega_\text{K}$                 & $\Omega_\text{K} \equiv \sqrt{GM/R^3}$                                                   & Orbital Frequency and Keplerian Frequency                    \\
$v_\text{K}$                                        & $v_\text{K} \equiv \sqrt{GM/R}$                                                          & Keplerian Velocity                                           \\
$c_\text{s}$                                        &                                                                                          & Sound Speed                                                  \\
$P$                                                 &                                                                                          & Gas Pressure                                                 \\
$E_\text{g}$                                        &                                                                                          & Gas Energy                                                   \\
$T_\text{s}$                                        &                                                                                          & Dust Stopping Time                                           \\
$St$                                                & $St \equiv \Omega_\text{K} T_\text{s}$                                                   & Dust Stokes Number                                           \\
$\Phi$                                              & $\Phi \equiv -GM/r$                                                                      & Stellar Gravity Potential                                    \\
$G$                                                 &                                                                                          & Gravitational Constant                                       \\
$M$                                                 &                                                                                          & Stellar Mass                                                 \\
$H_\text{g},\;H_\text{d}$                           & $H_\text{g} \equiv c_\text{s}/\Omega_\text{g}$, Equation~\ref{eq:dust_scaleheight}       & Scale Heights of Gas and Dust                                \\
$h$                                                 & $h \equiv H_\text{g}/R$                                                                       & Disk Aspect Ratio                                            \\
$p$                                                 & $p \equiv d\log\rho_\text{g,midplane}/d\log R$                                  & Power Law Index for Gas Density at the midplane                             \\
$q$                                                 & $q \equiv d\log c_\text{s}^2/d\log R$                                            & Power Law Index for Sound Speed Square                       \\
$\Sigma_\text{g},\;\Sigma_\text{d}$                 & $\Sigma \equiv \int^{z_\text{max}}_{z_\text{min}} \rho dz$                               & Surface Densities for Gas and Dust                           \\
$s$                                                 & $s \equiv p_\text{g}/\Sigma_\text{g}^{\gamma}$                                                & Specific Gas Entropy                                         \\
$\gamma$                                            &                                                                                          & Adiabatic Index                                              \\
$\rho_\text{eff}$                                   & $\rho_\text{eff} \equiv \rho_\text{d} + \rho_\text{g}$                                                          & Effective Density                                            \\
$Z_\text{d}$                                        & $Z_\text{d} \equiv \Sigma_\text{d,init}/\Sigma_\text{g,init}$                                      & Initial Surface Density Ratio Between Dust and Gas           \\
$\epsilon$                                          & $\epsilon \equiv \rho_\text{d}/\rho_\text{g}$                                            & Density Ratio Between Dust and Gas                           \\
$\eta$                                              & $\eta \equiv 1/2 (c_\text{s}/v_\text{K})^2 (d\log P/d\log R)$                                                          & Strength of Radial Pressure Gradient                     \\
$Ma_R,\;Ma_\phi,\;Ma_z$                             & Equation~\ref{eq:Ma}                                 & Mach Numbers in Cylindrical Coordinate                       \\
$\delta E_\text{kin}$                               & Equation~\ref{eq:kinerg}                                                                 & Total Kinetic Energy Density Variation                              \\
$\alpha_{R\phi}$                   & Equation~\ref{eq:alpha_Rphi}                                     & Effective Turbulent Viscosity Stresses   \\
$\mathscr{L}$                                       & Equation~\ref{eq:Lfunc}                                                                  & RWI Key Function (Entropy Modified Inverse Vortensity)       \\
$\mathscr{L}_\text{iso,midplane}$                   & Equation~\ref{eq:Lfunc_iso}                                                              & RWI Key Function at the midplane for Locally Isothermal Disks \\
$\omega_z$                 & Equation~\ref{eq:vorticity}                & Vertical Vorticity in Cylindrical Coordinate                    \\
$\chi$                                              & Equation~\ref{eq:aspect}                                                                 & Aspect Ratio of Vortex                                       \\
$S_\text{eff}$                                      & Equation~\ref{eq:entropy}                                                                & Effective Entropy                                            \\
$N_{z,\text{eff}}^2$                                & Equation~\ref{eq:buoyancy}                                                               & Effective Buoyancy Frequency Square                          \\
$Q$                                                 & $Q\equiv c_\text{s} \Omega_\text{K}/(\pi G \Sigma_\text{g})$                             & Toomre Q Parameter                                           \\
$\rho_\text{R}$                                     & Equation~\ref{eq:Roche}                                                                  & Roche Density                                                \\
$\epsilon_\text{crit}$                              & Equation~\ref{eq:crit_ratio}                                                             & Critical Dust-Gas Density Ratio at Roche Density             \\
$r_\text{c}$                                        &                                                                                          & Radial Center of Vortex in Spherical Polar Coordinate        \\
\bottomrule
\end{tabular}
\label{tab:symbols}
\end{footnotesize}
\end{table}

\section[]{Effects of Vertical Computational Domain on 2D VSI + SI Turbulence}\label{app:different_H}

\begin{figure*}[htp]
\centering
\includegraphics[scale=0.5]{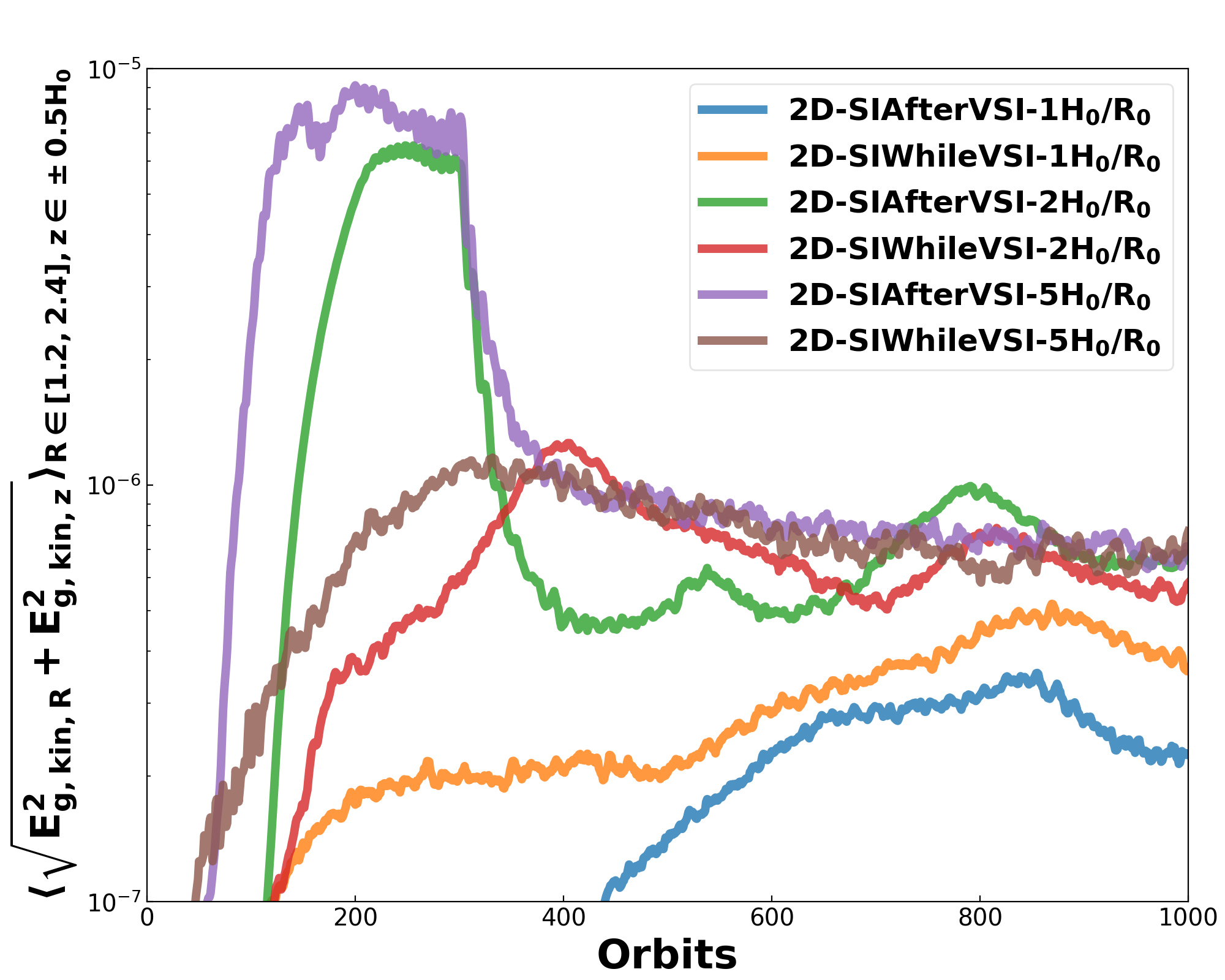}
\caption{The temporal evolution of the radially and vertically averaged gas kinetic energy, $\sqrt{E_{\text{g,kin},R}^2 + E_{\text{g,kin},z}^2}$, for different ``2D-SIAfterVSI'' and ``2D-SIWhileVSI'' models. Different values of $H_0/R_0$ indicate that different meridional domains are considered. ``2D-SIAfterVSI-5$H_0/R_0$'' and ``2D-SIWhileVSI-5$H_0/R_0$'' are the 2D VSI+SI models presented in the main text. All these models have the same refinement structures. The averaged domain covers $R \in [1.2, 2.4]$ and $z \in \pm 0.5 H_0$.}
\label{fig:different_H}
\end{figure*}

We performed a series of comparative simulations with varying vertical computational domains to assess their influence on the development of 2D VSI and SI turbulence. The tested meridional extents include $\theta \in \pm H_0/R_0$, $\theta \in \pm 2H_0/R_0$, and the default setup in the main text, $\theta \in \pm 5H_0/R_0$. These tests are denoted as $1H_0/R_0$, $2H_0/R_0$, and $5H_0/R_0$ in Figure~\ref{fig:different_H}. This figure shows the temporal evolution of the radially and vertically averaged gas kinetic energy (radial and vertical components), which serves as a proxy for the strength of the VSI+SI driven turbulence.

In the case with smallest vertical domain ($1\,H_0/R_0$), both the ``2D-SIAfterVSI-$1H_0/R_0$'' and ``2D-SIWhileVSI-$1H_0/R_0$'' cases exhibit significantly suppressed turbulence, maintaining low kinetic energy levels throughout the simulation. Notably, the ``2D-SIWhileVSI-$1H_0/R_0$'' case shows a higher turbulence level than the ``2D-SIAfterVSI-$1H_0/R_0$'' case, as SI-driven turbulence dominates in the former. However, VSI remains underdeveloped in both cases due to the limited vertical extent.

When the vertical domain is extended to $2\,H_0/R_0$, the ``2D-SIAfterVSI-$2H_0/R_0$'' case exhibits a rapid increase in kinetic energy before the dust is introduced, followed by a sharp decline due to dusty buoyancy (after 300 orbits). The turbulence level eventually reaches a value comparable to that of the ``2D-SIWhileVSI-$2H_0/R_0$'' case. Nevertheless, clear differences remain in the temporal evolution of kinetic energy between the two setups, suggesting that a vertical extent of $\pm 2H_0/R_0$ is still insufficient for the full development of VSI turbulence.

Most notably, in the full vertical domain case ($5\,H_0/R_0$), the turbulence levels in both ``2D-SIAfterVSI-$5H_0/R_0$'' and ``2D-SIWhileVSI-$5H_0/R_0$'' reach convergence relatively quickly (after approximately 400 orbits) and smoothly. This exemplifies that a sufficiently large vertical domain enables the full development and saturation of the VSI modes, which is also necessary to accurately capture the dynamics of VSI+SI driven turbulence.

\section[]{Axisymmetry of 3D VSI Turbulence}\label{app:axisymmetry}

\begin{figure*}[htp]
\centering
\includegraphics[scale=0.50]{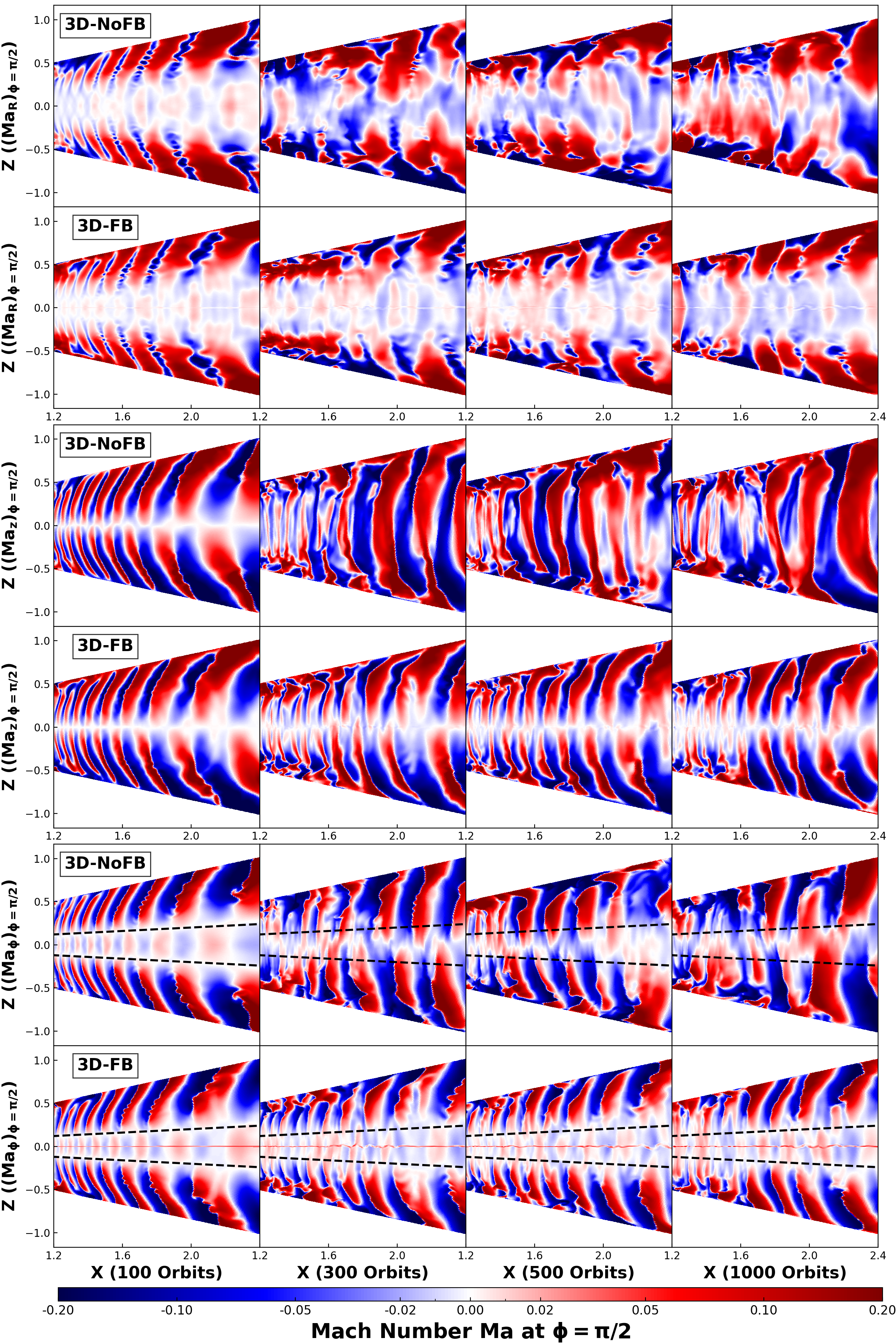}
\caption{Similar to Figure~\ref{fig:3D_Mach_numbers}, but for the Mach number $Ma$ at $\phi = \frac{\pi}{2}$, are depicted for 3D models at various time intervals.}
\label{fig:3D_Mach_numbers_section}
\end{figure*}

In the ``3D-NoFB'' and ``3D-FB'' models, we select the Mach numbers at the azimuthal angle \(\phi = \frac{\pi}{2}\) in Figure~\ref{fig:3D_Mach_numbers_section} to compare with the azimuthally averaged Mach numbers shown in Figure~\ref{fig:3D_Mach_numbers}. Since the mesh structures are axisymmetric and periodic boundary conditions are used in the azimuthal direction, the choice of azimuthal slice is arbitrary. We observe that Figure~\ref{fig:3D_Mach_numbers_section} exhibits similar patterns with minor fluctuations compared to Figure~\ref{fig:3D_Mach_numbers}, regardless of whether it is the ``3D-NoFB'' or ``3D-FB'' model. The non-axisymmetric instability (RWI) does not create significant azimuthal asymmetries in these two 3D runs.

\bibliographystyle{aasjournal}
\bibliography{references}{}

\end{document}